\documentclass[preprint,authoryear]{elsarticle}
\usepackage{amsmath,amssymb,mathabx,xfrac,siunitx,xfrac}
\usepackage{caption}
\usepackage{fancyhdr}
\usepackage{fullpage}
\usepackage{graphicx}
\usepackage{hyperref}
\usepackage{multirow}
\usepackage{pdflscape}
\usepackage{subcaption}

\journal{Icarus}

\biboptions{round,semicolon}

\begin{document}

\begin{frontmatter}

\title{Thermal Alteration of Labile Elements in Carbonaceous Chondrites}

\author[lpl]{Alessondra Springmann\corref{cor1}}    
\ead{sondy@lpl.arizona.edu}

\author[lpl]{Dante S.~Lauretta}
\author[tsi]{Bjoern Klaue} 
\author[jpl]{Yulia S.~Goreva}
\author[um]{Joel D.~Blum}
\author[cze]{Alexandre Andronikov}
\author[ut,psi]{Jordan K.~Steckloff}
    
\cortext[cor1]{Corresponding author} 

\address[lpl]{Lunar and Planetary Laboratory, University of Arizona, Tucson, AZ 85721, United States} 
\address[tsi]{TSI GMBH, Neuk\"{o}llner Strasse 4, 52068 Aachen, Germany}
\address[jpl]{NASA Jet Propulsion Laboratory, Pasadena, CA 91109, United States}
\address[um]{University of Michigan, Department of Earth and Environmental Sciences, Ann Arbor, MI 48109, United States}
\address[cze]{Czech Geological Survey, Geologicka 6, 152 00 Prague, Czech Republic}
\address[ut]{Department of Aerospace Engineering and Engineering Mechanics, University of Texas at Austin, Austin, TX 78712, United States}
\address[psi]{Planetary Science Institute, Tucson, AZ 85719, United States}

\begin{abstract}
Carbonaceous chondrite meteorites are some of the oldest Solar System planetary materials available for study. The CI group has bulk abundances of elements similar to those of the solar photosphere. Of particular interest in carbonaceous chondrite compositions are labile elements, which vaporize and mobilize efficiently during post-accretionary parent-body heating events. Thus, they can record low-temperature alteration events throughout asteroid evolution. However, the precise nature of labile-element mobilization in planetary materials is unknown. Here we characterize the thermally induced movements of the labile elements S, As, Se, Te, Cd, Sb, and Hg in carbonaceous chondrites by conducting experimental simulations of volatile-element mobilization during thermal metamorphism. This process results in appreciable loss of some elements at temperatures as low as 500 K. This work builds on previous laboratory heating experiments on primitive meteorites and shows the sensitivity of chondrite compositions to excursions in temperature. Elements such as S and Hg have the most active response to temperature across different meteorite groups. Labile element mobilization in primitive meteorites is essential for quantifying elemental fractionation that occurred on asteroids early in Solar System history. 
This work is relevant to maintaining a pristine sample from asteroid (101955) Bennu from the OSIRIS-REx mission and constraining the past orbital history of Bennu.  Additionally, we discuss thermal effects on surface processes of near-Earth asteroids, including the thermal history of ``rock comets'' such as (3200) Phaethon. This work is also critical for constraining the concentrations of contaminants in vaporized water extracted from asteroid regolith as part of future \textit{in situ} resource utilization for sustained robotic and human space exploration. 
\end{abstract}


\end{frontmatter}


\section{Introduction}
Carbonaceous chondrites are meteorites assembled from some of the oldest and most volatile Solar System materials. The CI chondrites have bulk abundances for most elements similar to those of the solar photosphere \citep{Anders1989}; exceptions are C, N, O, and the noble gases. The bulk compositions of other types of carbonaceous chondrites such as the CM, CO, and CV groups show depletions in many volatile elements relative to CI chondrites. 

Cosmochemists divide volatile elements into two categories: moderately volatile and highly volatile. This distinction is based on equilibrium chemistry predicted by thermodynamic modeling of condensation in the primordial solar nebula \citep{Larimer1967}.  Moderately volatile elements have condensation temperatures between those of Mg-silicates and FeS. Highly volatile elements have condensation temperatures below that of FeS. The depletion of volatiles is one of the essential fractionation processes affecting solid material in the Solar System \citep{Lipschutz1988}.  This process is mainly responsible for determining the bulk volatile-element abundances in Earth and other terrestrial planets. 

The two main competing theories to explain volatile depletions in carbonaceous chondrites are the condensation model \citep{Wasson1974,Wai1977} and the two-component model \citep{Wood1963,Anders1964,Larimer1967}.  
In the former model, continuous loss of nebular gas from the chondrite-forming region during condensation results in the observed elemental depletions. 
In the latter model, chondrites are mixtures of coarse-grained refractory chondrules depleted in volatiles and fine-grained volatile-rich CI chondritic matrix \citep{Alexander2001}. 
Another possibility is volatile loss occurred during processes related to post-accretionary heating of the near solar-composition material. Such processes include aqueous alteration, metamorphic heating, and impact shock. Certain volatile elements, called the labile elements, are highly mobile during such processing on asteroids \citep{Lipschutz1988}.  Thus, distributions of volatile trace elements in meteorites can provide information about the intensity and duration of these events.

A large number of experimental simulations study volatile-element loss during thermal metamorphism (Table \ref{tab:metamorphism}); however, little is known about volatile-element host phases within meteorites. Furthermore, the details of their thermal release profiles have not been determined because investigations into the effect of thermal metamorphism on these elements involved bulk analyses of experimentally heated samples \citep{Xiao1992}. Such details could provide valuable information about trace-element host phases and distributions within chondritic meteorites. If trace elements are contained within a single host phase, the heating of the meteoritic material in an open system should result in the loss of a volatile component over a narrow temperature range \citep{Anders1977}. Alternatively, volatile trace elements may be present in a variety of phases, and a proportionate amount of each element could be lost from each host gradually by diffusion or suddenly during a phase change \citep{Wood1967,Dodd1969,Wasson1972}.

We developed a technique to simultaneously measure the thermal release profiles of a suite of labile elements by coupling a programmable tube furnace to a single-collector magnetic sector inductively coupled plasma mass spectrometer (ICP-MS). This method allows for determination of element release as a function of temperature \citep{Lynam2013}. This technique was first used in our measurements of the isotopic composition of Hg in meteorites \citep{Lauretta2001}.  We performed thermal release experiments in an attempt to isolate isotopically anomalous Hg. No anomalies were found, but we did notice differences in the Hg chemistry of CV and CM chondrites. We decided to expand our study to include other highly and moderately volatile elements and additional meteorite groups. This study builds on the work of previous researchers by providing information about the temperatures when these elements are released, correlations among the release patterns of the different elements, and the quantification of the abundance of material released from the samples.

We consider the applications of laboratory experiments tracking thermal release of labile elements with temperature, in particular to near-Earth asteroid surface studies.  In particular, we discuss the effects of temperature on maintaining a pristine sample to be returned from OSIRIS-REx mission target, asteroid (101955) Bennu, as well as constraining the the past orbital history of Bennu (Section \ref{sec:orex}).  Additionally, we examine thermal surface processes on near-Earth asteroids including (3200) Phaethon (Section \ref{sec:phaethon}), and also present the concentrations of contaminants in vaporized water extracted from asteroid regolith, relevant for future \textit{in situ} resource utilization for sustained robotic or human space exploration (Section \ref{sec:isru}). 

\section{Methods}

\subsection{Thermodynamic Calculations}
Thermodynamic calculations were performed using the equilibrium module in the HSC Chemistry 5.1 software package, produced by Outokumpu Research Oy. This module calculates multi-component equilibrium compositions in heterogeneous systems by using a database of more than 15,000 compounds and the GIBBS or SOLGASMIX Gibbs energy minimization equilibrium solvers \citep{White1958,Eriksson1990}. We determined the equilibrium distribution of H, He, C, N, O, Mg, Si, Fe, S, Cl, As, Se, Cd, Sb, and Te among more than 280 gas phase species, as well as chlorides, silicates, oxides, sulfates, sulfides, nitrides, carbides, and metals under conditions relevant to the solar nebula.  The total pressure was 10$^{-4}$ bars and the temperature range considered was 200 to 2000 K.

\subsection{Sample Preparation}
All meteorite samples were obtained from the Center for Meteorite Studies at Arizona State University. Clean, interior samples of the CI chondrite Orgueil; the CM chondrites Murchison, Murray, Nogoya, and Cold Bokkeveld; the CO chondrites Kainsaz, Ornans, and Isna; and the CV chondrites Allende, Vigarano, Mokoia, and Grosnaja were obtained by grinding away fusion crust with a high-speed diamond sanding wheel and immediately isolating interior pieces. These pieces were disaggregated under clean-room conditions to obtain uncontaminated material from the center. This material was then gently crushed under ethanol to obtain visually uniformly sized powders. Three different samples of Murchison and Allende were obtained to check for variability across individual specimens. 

\subsection{Experimental Techniques}
\subsubsection{Volatile Release During Heating}
We determined thermal release profiles (Figure \ref{fig:abundances}) online by coupling a programmable tube furnace to an ELEMENT 2 single-collector magnetic-sector inductively coupled plasma mass spectrometer (ICP-MS).  The furnace consisted of a 22-mm (inside diameter) thin-wall, quartz-glass tube wrapped with high-temperature, ceramic-beaded Ni-ChromeV resistance wire over a length of $\sim10$ cm. In the middle of the heated area a 6-mm ID quartz glass thermowell was inserted holding a type-K thermocouple. Temperature control and data logging were performed with a Digi-Sense Advanced Temperature Controller (Cole Parmer, Vernon Hills, IL, USA). The temperature probe was calibrated using the ``Sensor Calibrate'' function of the temperature controller. This function enables the operator to calibrate out sensor error to give a more accurate reading. Per the manufacturer guidelines, the calibration was done using both an ice bath and boiling water by adjusting the offset until 273 K and 373 K were displayed, respectively.  The accuracy of the type-K thermocouple is $\pm 0.1$\%, or $\pm 0.4$ K, according to manufacturer specifications.

Small (1--100 mg) aliquots of powdered meteorites and standards were placed in glazed porcelain boats ($5\times7$ mm) in the isothermal region of the furnace. The samples were heated from room temperature to 1173 K along a linear temperature ramp over 25 minutes.  This method allowed for determining of when distinct phases were releasing each element. Because of the high sensitivity of the ELEMENT 2 ICP-MS and the low background levels, we achieved detection limits of less than 1 fg. None of the thermal decomposition procedures published previously for meteorites provided this information. 

In our first set of experiments we monitored the release of a large suite of volatile siderophile and chalcophile elements (Zn, As, Se, Cd, In, Sn, Sb, Te, Pt, Hg, Au, Tl, Pb, and Bi). However, we noticed that many of these elements were not released in detectable quantities until the highest temperatures in our experiments. A subset of these elements---As, Se, Te, Cd, Sb, and Hg---became liberated from the samples at much lower temperatures. We focus on these elements to gain insight into element redistribution during low-temperature alteration of carbonaceous chondrites. In these experiments we used low mass-resolution mode (M/$\Delta$M(5\%)=400) to measure the abundance of a single isotope for each of these elements ($^{34}$S, $^{75}$As, $^{77}$Se, $^{111}$Cd, $^{121}$Sb, $^{125}$Te, and $^{202}$Hg). 

We introduced yttrium (Y) as an internal standard to allow direct comparison of the amount of each element released from the meteorites. A membrane desolvation unit (ARIDUS, CETAC) was used to provide a dry Y-nitrate aerosol, which we mixed with the sample gas, Ar, before introduction to the tube furnace. The abundance of Y introduced into the plasma remained constant throughout the experiment. As an initial estimate, the measured ion counts were corrected for each element to account for its isotopic abundance and ionization efficiency in the plasma \citep{Houk1986}. We then applied the absolute abundance measurements to quantify the amount and percentage of each element lost as a function of temperature.  Argon was used to make an inert atmosphere when the samples were heated, with a flow rate of 1 L/minute.  

Thermal release profiles could be influenced by using crushed samples with variable grain sizes.  For example, diffusive loss of elements from host phases and/or within a particle could be one of the controlling factors of sublimation of elements from the samples. In addition, the ramp rate could have substantial effects on the sublimation behavior of labile elements during temperature increase. The thermal release profile was varied in early pathfinder experiments. The identification of distinct peaks in the chosen, 25-minute ramp indicated the carrier phases for the elements of interest were undergoing thermal decomposition and releasing their trapped labile elements. Thus, we do not consider diffusion to be the prime mechanism for element loss when distinct peaks are present.  When a linear heating technique produces a sharp peak in element release, it is related to a phase transformation of the carrier of the relevant elements.  In these cases, grain sizes and diffusion do not control the release profile of the elements of interest \citep{Boynton2001}.  After each experiment, the furnace was heated to peak temperatures to determine if there were any re-condensed species.  In all cases, there was no observable re-release of elements.  The 1-L/minute flow rate of Ar gas prevented re-condensation of species downstream. 

\subsubsection{Element Quantification}
We determined the total amount of each element released during the heating experiment by measuring the bulk abundances in acid-digested portions of the unheated meteorite samples, as well as the heated samples.  We measured bulk abundances in unheated samples and found they were consistent with published compositional data; thus, we did not lose labile elements during sample acid dissolution.  For these bulk abundance measurements, 7 to 50 mg of the pulverized material from each sample was digested in a concentrated HNO$_3$-HF mixture with the addition of a few drops of HClO$_4$ in 15 mL Teflon beakers on a hot plate at 400 K for 48 hours. After cooling, we evaporated this solution and then reconstituted it with 5\% HNO$_3$ before introduction to the ICP-MS. The analytic solutions contained $\sim$100 $\mu$g/mL of total dissolved solids. A blank analytical solution was prepared using the same procedure. Solution standards consisting of known amounts of the analyzed elements were developed using single-element solutions (High-Purity Standards; Charleston, SC). A standard-blank solution was prepared at the same time using successive dilutions of the 5\% HNO$_3$ standard-carrier solution. Sample concentrations were determined by first subtracting blank signal intensities from those obtained from the sample and standard solutions. By identifying the change in the bulk abundance of every element lost from the heated samples, we were able to calculate the percentage lost as a function of temperature (Figure \ref{fig:percentlost}).  The heating experiments consumed the entire available sample mass of Kainsaz (CO); the total elemental loss as a result of heating was not measured for this meteorite.

\section{Results}

\subsection{Condensation Chemistry in the Solar Nebula}
\begin{figure}[]
\centering
\includegraphics[width=0.36\columnwidth]{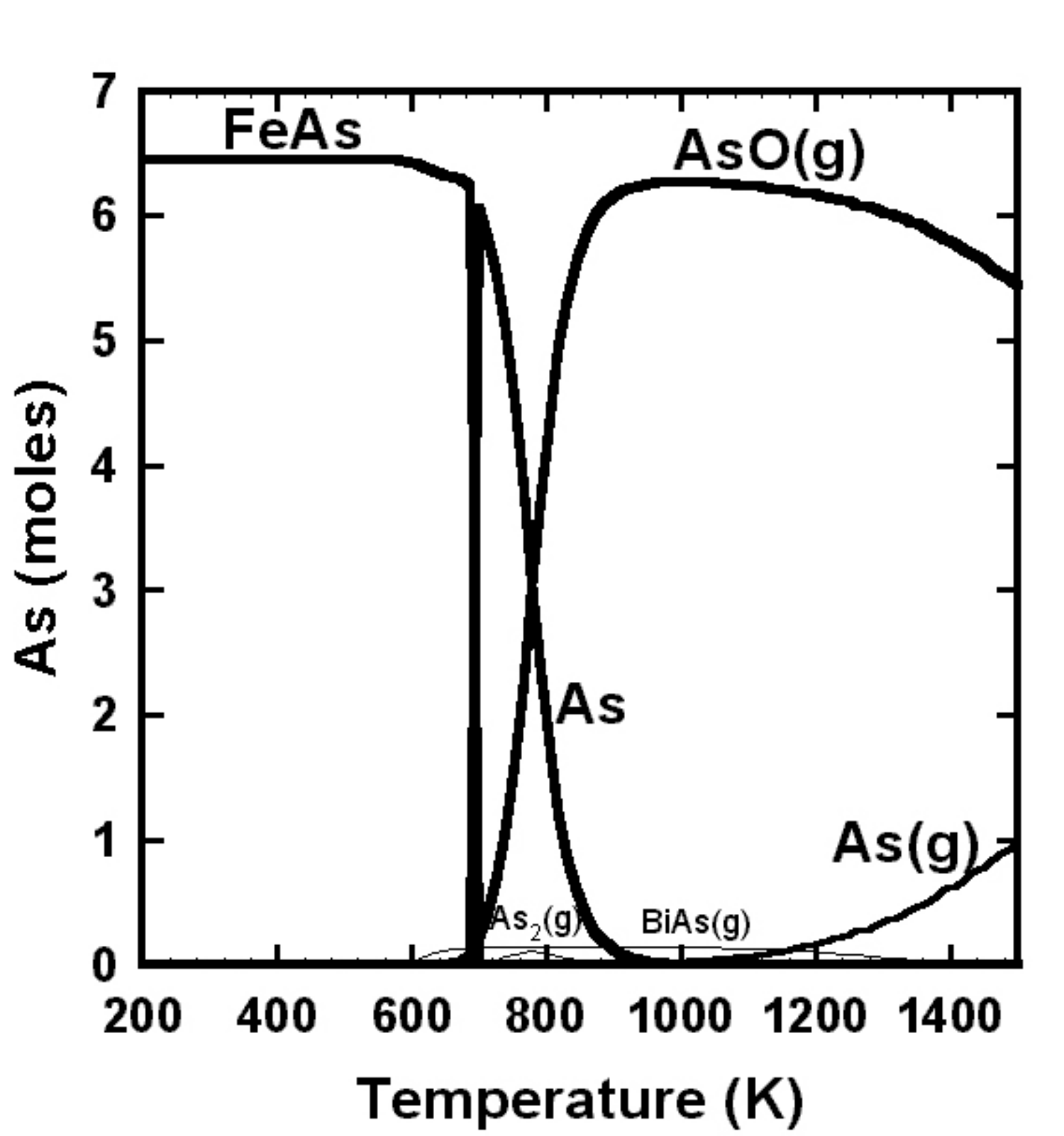}
\includegraphics[width=0.37\columnwidth]{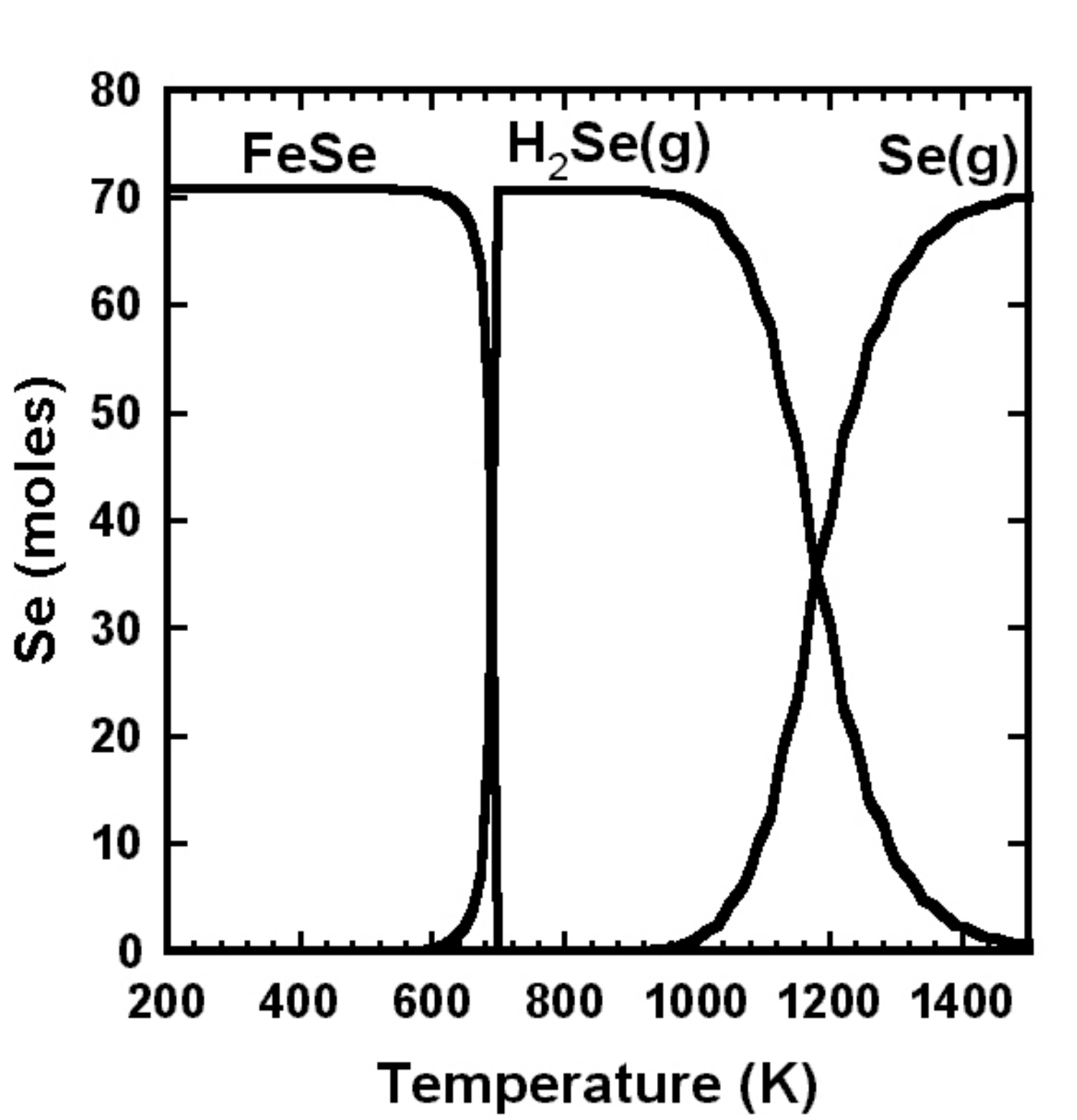}\\
\includegraphics[width=0.36\columnwidth]{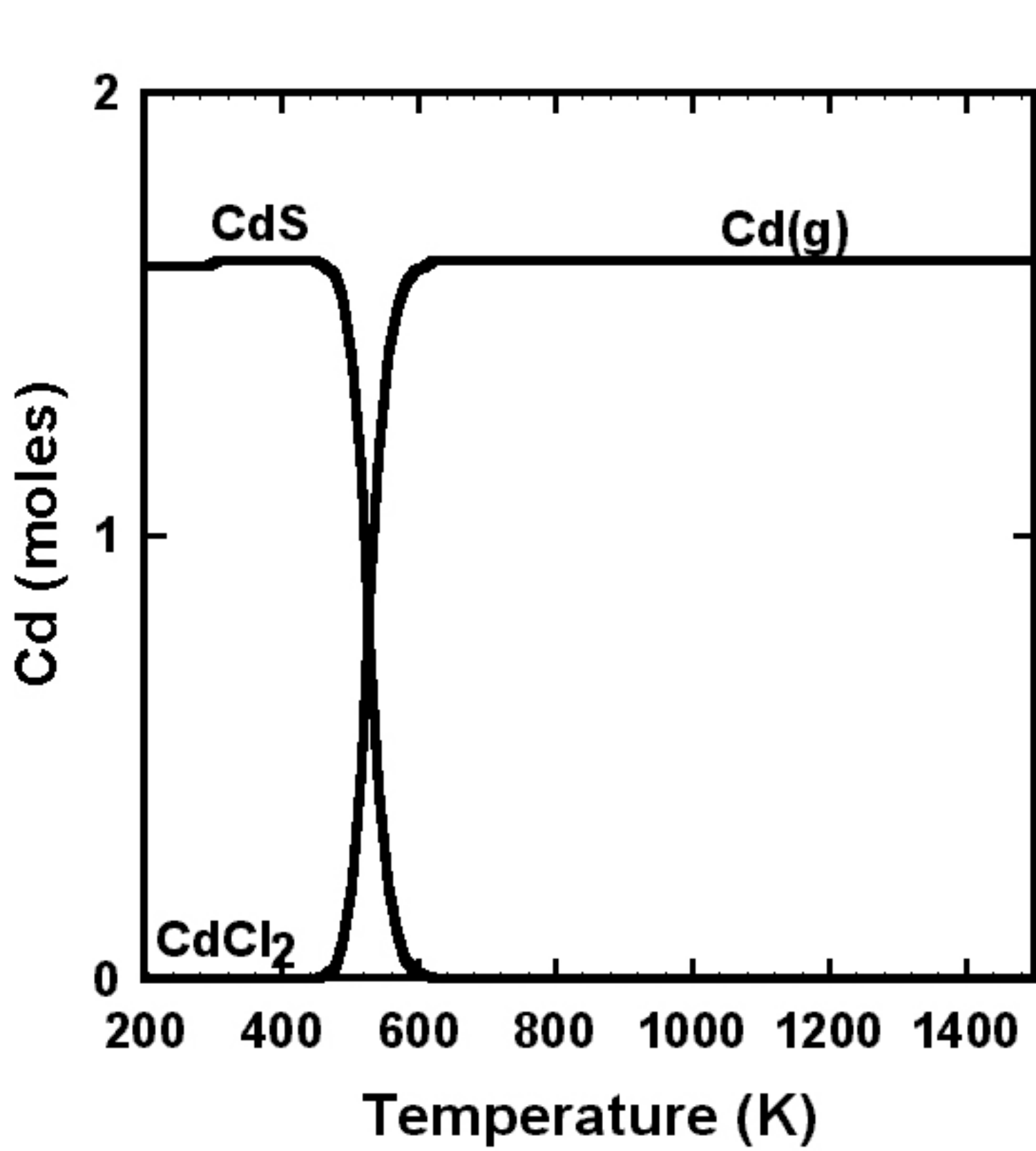}
\includegraphics[width=0.40\columnwidth]{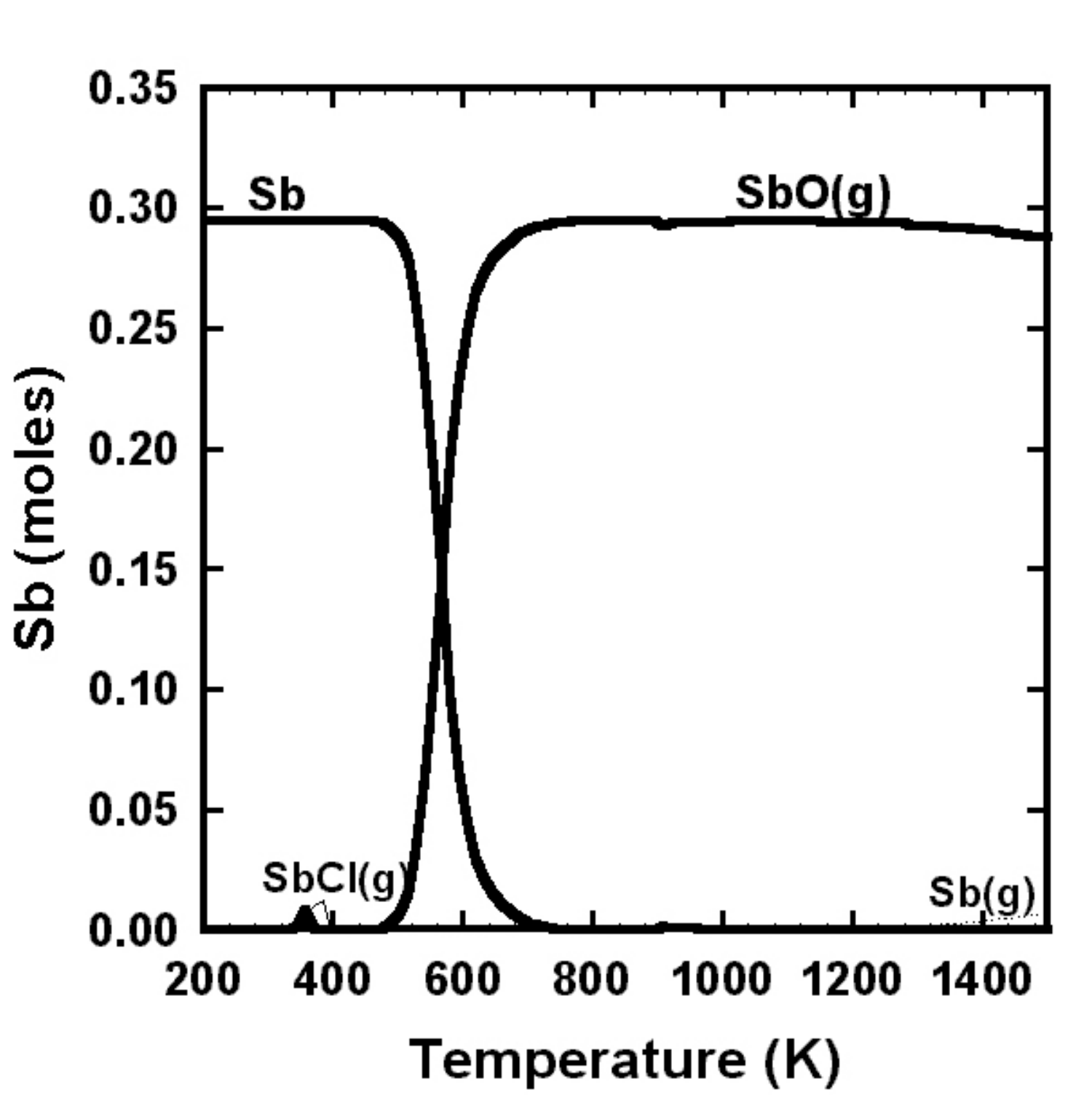}\\
\includegraphics[width=0.36\columnwidth]{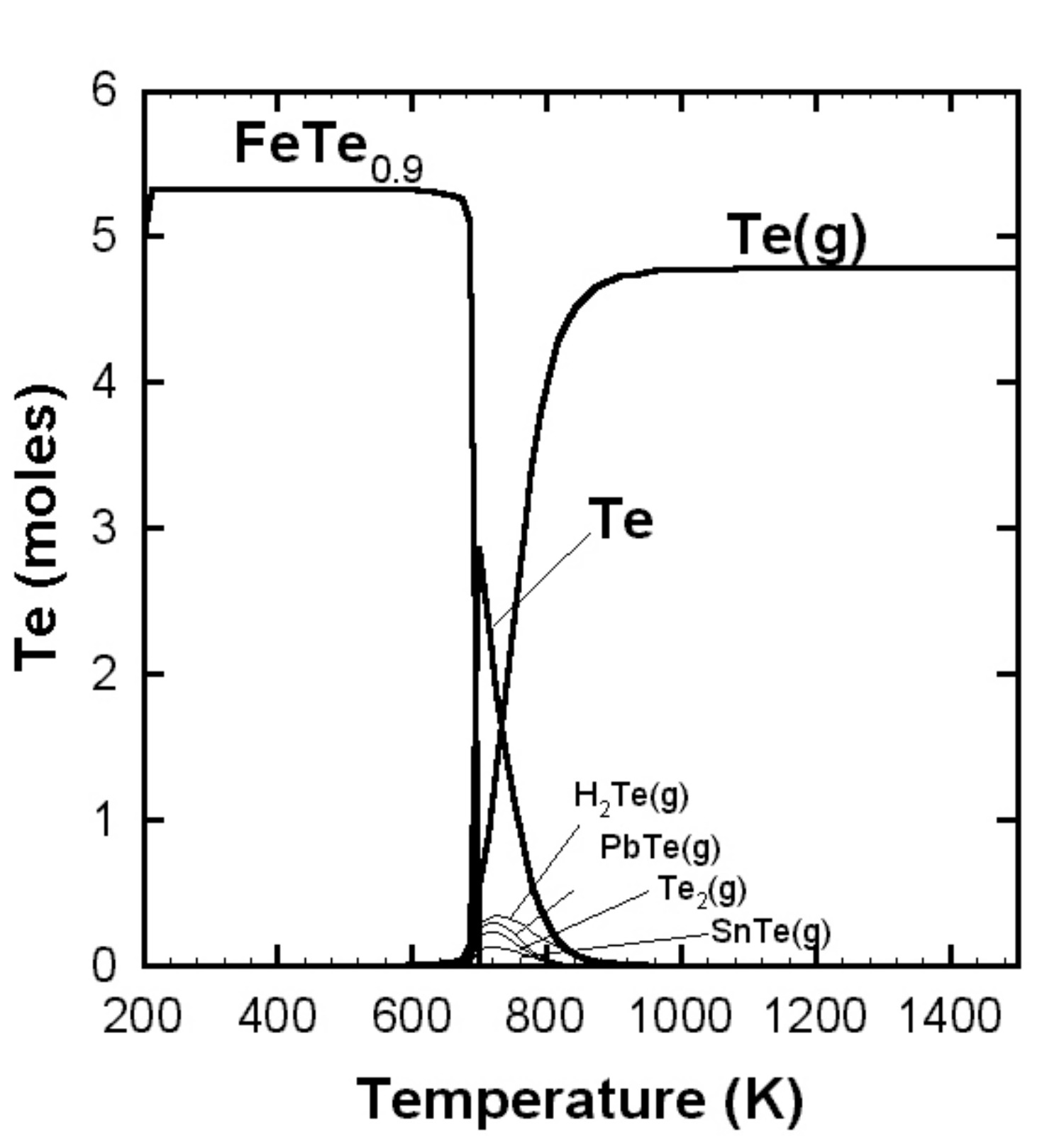}
\caption{Equilibrium distribution calculations for elements As, Se, Cd, Sb, and Te showing gas and mineral phases likely to form under conditions relevant to the solar nebula (total pressure 10$^{-4}$ bars; temperature range 200--1600 K).}
\label{fig:thermocalc}
\end{figure}

We determined the equilibrium distribution of As, Se, Te, Cd, and Sb in the solar nebula at a total pressure of 10$^{-4}$ bars (Figure \ref{fig:thermocalc}).  All of these elements are predicted to condense into metal or sulfides. The 50\% condensation temperatures we determined are consistent with prior studies of these elements \citep[e.g.][]{Lodders2003}. We found As and Te initially condense as trace elements in the bulk Fe-based alloy but transfer to the sulfide phase as the components FeAs and FeTe once metal sulfidation begins. Selenium and Cd both condense directly into the sulfide phase, with Se condensing as FeSe and Cd as CdS in solid solution with troilite. Antimony is predicted to condense into the Fe alloy; however, its condensation temperature is well below that of sulfide formation. The mechanism by which Sb incorporates into the metal phase is unclear, because post-sulfidation, the metal will be covered by the growing sulfide layer. \citet{Lauretta1999} discuss the nebular chemistry of Hg, suggesting that it is incorporated into solid material in the early Solar System through chemisorption on metal or sulfide grains. We expect that nebular metals and sulfides are the original carriers of these elements on meteorite parent bodies. Any heating or alteration of these phases will result in the redistribution of these elements into low-temperature components.

\subsection{CI Chondrites}
Orgueil was the only CI chondrite studied.  All monitored elements released from the sample during heating (Figure \ref{fig:ab-orgueil}).  This meteorite released $\sim20$\% of its total S over the entire temperature range, with three distinct maxima in the S release profile at 530, 765, and 940 K.  The first peak falls between the two-peak release profile of Hg (470 and 575 K), perhaps owing to decomposition of troilite.  Hg release slowed between 750 and 900 K, increasing slowly above 900 K.   Other element release profiles have peaks within 35 K of the first S peak (As, 550 K; Se, 510 K; Sb, 565 K).  Se released continuously above 835 K, Te above 650 K.  

\subsection{CM Chondrites}
We studied the Murchison, Murray, Nogoya, and Cold Bokkeveld CM chondrites (Figures \ref{fig:ab-murchison}, \ref{fig:ab-murray}, \ref{fig:ab-nogoya}, and \ref{fig:abuncoldbok}). We chose these meteorites because they have experienced different degrees of aqueous alteration. Murchison has experienced the least amount of aqueous activity; Nogoya has experienced the most \citep{Browning1996}.

There are similarities among the release profiles across the four CM chondrites; however, the non-detection of S, Se, Te, and Hg for Murray reduces the potential for comparison.  
Murchison, Nogoya, and Cold Bokkeveld released detectable S with peaks at 625, 520, and 560 K and continued releasing measurably throughout the experiments.  
Arsenic release from Nogoya and Cold Bokkeveld increased above 400 K; in these two meteorites, As and Se have similar release profiles, though Nogoya shows a release profile with three peaks for Se.  Murchison and Murray have similar As release profiles.  
Sb has peaks at 640 and 650 K in Nogoya and Cold Bokkeveld, respectively.  Te release profiles for Nogoya and Cold Bokkeveld are similarly linear.  Murchison and Murray show similar release profiles for Sb.  

Detectable quantities of Hg were released from Murchison, Nogoya, and Cold Bokkeveld, with peaks at 645  and between 460 and 500 K for Murchison and Cold Bokkeveld, respectively.  Hg release from Cold Bokkeveld saturated the ICP-MS detector, represented by a dashed line (Figure \ref{fig:abuncoldbok}).  
The rate of Hg release decreased significantly above this temperature for Murchison and Cold Bokkeveld; Nogoya shows three peaks at 590, 740, and 910 K, continuing to release appreciable quantities above 450 K, with only a slight decrease above 910 K. 

\subsection{CV Chondrites}
We performed thermal release experiments with four different CV chondrites. Allende, Grosnaja, and Mokoia (Figures \ref{fig:ab-allende}, \ref{fig:ab-grosnaja}, and \ref{fig:ab-mokoia})  are members of the oxidized subgroup of the CV chondrites. Vigarano (Figure \ref{fig:ab-vigarano}) is also the only member of the reduced CV chondrite group analyzed in this study.

Sulfur release from Allende and Mokoia increased throughout the experiment. Mokoia shows more peaks in its release profile than Allende, with a large peak at 1035 K.  Sulfur release from Grosnaja was sufficient to saturate the ICP-MS detector between 700 and 980 K. The S release profile for Vigarano shows a peak at 615 K.  

The As and Se release profiles for Allende appear similar, with peaks at 870 and 1010 K; Grosnaja also has similar As and Se release profiles with peaks at 380 and 885 K.  Mokoia has a peak in As at 635 K, just above a peak in Se at 615 K.  Vigarano has corresponding peaks in As and Se at 675 K.  

Cd and Sb release in Allende peak at 1025 K (along with a peak in Te), and Sb has an additional peak at 880 K.  Release profiles of Cd, Sb, and Te follow a similar trend for Vigarano.  Mokoia and Vigarano have similar Cd and Sb follow similar release profiles, resembling the release of Cd from Grosnaja.  Sb has two peaks in the release profile for Grosnaja: at 380 and between 760 and 840 K.  

These four CV chondrites all show Hg release between 400 and 700 K.  During the experiment with Allende, Hg was the first element to be liberated from the meteorite. The Hg release profile contains a peak at 495 K, with a slow decrease in counts until 650 K.  Similarly, Mokoia has a large peak in its Hg release profile between 400 and 700 K, with a maximum at 565 K, close to a peak in Te release.  
Hg release from Grosnaja peaks at 535 K.  The Hg-release profile for Vigarano shows increased counts between 500 and 700 K, with peaks at 560 and 640 K.

\subsection{CO Chondrites}
We studied the three CO chondrites Kainsaz (CO3.2, Figure \ref{fig:ab-kainsaz}), Ornans (CO3.4, \ref{fig:abunornans}), and Isna (CO3.8, \ref{fig:abunisna}); all three released detectable amounts of the seven labile elements in this study.  For Kainsaz, As, Se, Cd, Sb, and Te all increased above 900 K; for Ornans and Isna, S, As, Se, Cd, Sb, and Te increased above 850 K.  

Kainsaz has two small peaks in S release at 465 and 655 K and a larger one at 870 K.  Ornans has a S peak at 520 K, close to peaks in As and Te, and after the release of the majority of Hg.  Sulfur counts for Isna increase above 600 K with a peak at 910 K.  Sulfur release in Isna follows similar profiles to those of Cd, Sb, and Te.  

Arsenic release for Kainsaz peaks at 530, 655, and 835 K, then increases in counts above 825 K.  Ornans has a large peak in As release at 550 K, increasing above 775 K.  Arsenic release from Isna is flat until it increases above 540 K, with a small peak at 910 K.  

Selenium has a peak at 645 K for Kainsaz, then increases above 1000 K.  Similarly, Cd and Sb also increase in counts above 1000 K for Kainsaz.   In Ornans, Se initially shows a wide peak between 310 and 530 K, then increases from 695 K.  Isna has small peaks in Se at 640 and 870 K, with increasing counts above 910 K.  

Cadmium shows a flat release profile for Kainsaz until 650 K, then sharply increases in counts at 930 K.  For Ornans, Cd release peaks at 920 K and continues increasing above 975 K.  The Cd release profile for Isna shows an increase above 925 K.  

The Sb release profile from Kainsaz initially decreases, then ramps up above 900 K, a similar profile to that from Ornans.  Antimony release for Isna initially is flat then increases above 800 K.  Tellurium has a peak at 625 K, then increases above 740 K.  For Ornans, Te has peaks at 545 and 1070 K.  Te is mostly flat from Isna and increases in rate above 945 K.  

From Kainsaz, Hg release peaks at 535 K; high counts from of Hg from Ornans saturated the detector, peaking between 420 and 485 K; and Isna shows a peak in Hg at 470 K.  

\section{Discussion}
\subsection{Element Distributions Resulting from Nebular Condensation}
In the process of nebular condensation, labile elements condense into metal and sulfide phases (Figure \ref{fig:thermocalc}). 
As a result, troilite (FeS) is the most abundant sulfide phase in many types of chondritic meteorites.  
Thermodynamic calculations suggest that heating of FeS under an inert gas will result in the establishment of a substantial equilibrium vapor pressure of S$_{2}$ at 1000 K.  
The equilibrium vapor pressure will increase to temperatures well above 1200 K, at which point the FeS phase decomposes \citep{Lauretta1997}; this result is consistent with previous experiments where sulfide loss was observed from samples of Allende and troilite heated to 1000 K and higher \citep{Ikramuddin1975gca,Tachibana1998}.

Type-3 CO chondrites are thought to be among the most primitive meteorites \citep{Alexander1989,Grossman1999}. Because the heating experiments consumed the entire available sample mass, the total elemental loss as a result of heating was not measured for Kainsaz; thus, we could not calculate the percentages of elements lost.  Ornans lost more than 95\% of its Hg, saturating the ICP-MS detector in the process; it lost between 5 and 15\% of the other six elements.  Isna lost 50\% of its Hg, more than 70\% its of Cd, and $<$50\% of its S and Sb.  
Element release from these samples, with the exception of Hg in all three and S in Kainsaz, tends to increase linearly with temperature (Figure \ref{fig:abunornans}, \ref{fig:abunisna}). This result indicates a more refractory phase, stable above 1200 K, contains the majority of trace elements. This phase is most likely troilite, which is abundant in CO chondrites. Because this phase is not decomposing, trace elements would migrate from the interior of the sample to the surface, most likely through diffusion. This behavior is very different from that of the other carbonaceous chondrites, indicating that low-temperature sulfide phases are not abundant in these meteorites. This observation is consistent with these meteorites experiencing minimal low-temperature alteration on their parent bodies. Thus, they likely preserve information about the condensation and distribution of labile elements before aqueous processes.  

\subsection{Element Distributions Resulting from Parent-Body Aqueous Alteration}
Labile element distributions are affected by parent-body aqueous alteration. The CI chondrites are thought to be the most chemically primitive meteorites available for study because their bulk abundances closely match those of the solar photosphere. However, these meteorites have experienced extensive aqueous processing on their parent asteroid, considerably altering their mineralogy. The CM group of carbonaceous chondrites also provides evidence of widespread aqueous alteration in the early Solar System. Petrologic work on these meteorites showed this class of chondrites represents an alteration sequence, with some members experiencing minimal fluid activity and others seeing intense alteration \citep{Zolensky1997}. Extensive alteration of metal is apparent in these meteorites, and metal grains are surrounded by the clay-like minerals tochilinite [(FeS)$\cdot$Fe(OH)$_2$] and cronstedtite [(Fe$^{+2}$,Fe$^{+3}$)$_3$(Fe$^{+3}$,Si)$_2$O$_5$(OH)$_4$] \citep{Tomeoka1989}.The extent of alteration has been quantified in a mineralogical alteration index, based on the Fe and Mg content of the matrix phyllosilicates  \citep{Browning1996}. Recent studies have shown that this index also correlates with other indicators of aqueous activity such as the stable isotopic composition of carbonates \citep{Benedix2000}. 

We observed appreciable S release from the CI and CM chondrites well below the temperature of troilite decomposition. For many samples, S, As, Se, Sb, Te, Cd, and Hg were released at detectable levels when the meteorites were heated to 1100 K. Maxima occur at relatively low temperatures in the release profiles of As, Se, Sb, and Te. In many cases, the release of these elements correlates with maxima in the release of S. A low-temperature sulfide phase could be present in many of these meteorites, serving as a host phase for these trace elements. The relative abundances of As, Se, Sb, and Te in this potential low-temperature phase are highly variable.

Orgueil released the highest amounts of S, Se, and Te of any of the carbonaceous chondrites studied.  The release of these three elements was correlated; all of them had maxima at 550 K. Elemental S will decompose by 625 K under an inert atmosphere. This phase is known to occur in Orgueil and may be responsible for the low-temperature peak in the S release profile. Orgueil also released a relatively large quantity of Cd, with a maximum at 1000 K. This behavior is consistent with the known history of this meteorite, which experienced intensive aqueous alteration at temperatures above 375 K. It is surprising that very little As and Sb were released from Orgueil during heating. This result indicates that these elements were not very mobile during alteration on the Orgueil parent body.

The elements As, Se, and Te were released simultaneously from the four CM chondrites that were studied. The ratios of these elements to S (normalized to CI chondrite abundances) correlate with the mineralogy of the host meteorites. Release from CM chondrites between 725 and 875 K increases with the degree of aqueous alteration (Cold Bokkeveld, Nogoya, Murray, and Murchison), suggesting that these elements mobilized on the parent body. This sequence correlates with the degree of aqueous alteration in the CM chondrites. Such a correlation suggests that these elements may also be tracking secondary alteration processes on asteroidal bodies. Because these labile elements are released from the meteorites at temperatures well below the decomposition temperature for troilite, they must have transferred to a low-temperature S-bearing phase. Furthermore, as their abundance relative to S increases with increasing alteration in the CM chondrites, the host phase must also increase.

\subsection{Element Distributions Resulting from Thermal Metamorphism}
The loss patterns of labile elements in the CV chondrites analyzed reflect the thermal processes experienced by parent bodies of these meteorites. In all of the CV chondrite samples examined, there are maxima in the release profiles of many elements at 700 K. The abundance of the different volatile elements released at this temperature varies among the different samples analyzed. Allende and Grosnaja released small amounts of As and Se.  Mokoia released more substantial amounts of As and Se and lesser Cd at 700 K. In contrast, Vigarano, which is a member of the reduced CV subgroup, released substantial Te in addition to As, Se, and S at 700 K. Overall, Vigarano releases less volatile elements than the oxidized CV meteorites. These results suggest that the volatile trace elements are contained in more refractory phases in Vigarano and more volatile phases in the other CV chondrites. This dichotomy is consistent with the theory of oxidized CV chondrites experiencing considerable aqueous alteration followed by dehydration. 
Elements released from the CV meteorites below 725 K most likely recondensed after metamorphism.

Our experiments provide information about the rate of volatile loss from carbonaceous asteroids during thermal metamorphism. These experiments simulate the loss of elements during heating in a vacuum. Using these results, we calculate the rate of volatile loss during heating.  We determine these values at discrete temperature intervals between 400 and 1200 K. Thermally labile elements release in distinct peaks at low temperatures (below 1200 K). The order in which these elements are lost from carbonaceous chondrites changes with temperature. For example, a sample of Allende heated at 475 K would lose all of its S in 57 days but would require almost 9000 days to lose its Cd. At 1100 K the same sample of Allende would degas all of its Cd in 0.18 days but would take more than two days to lose all of its S.

The rate of volatile loss from these samples is not related to the condensation sequence in the solar nebula.  Instead, many factors such as element redistribution during parent-body processes and the oxidation state of the atmosphere during heating likely play an important role in the rate of volatile loss. Further, the kinetics of volatile degassing are very rapid, compared to the timescales of secondary alteration on planetesimals. Thus, it seems that extensive volatile loss did not occur as a result of parent body alteration, and that volatiles redistributed during these events.  

\subsection{Numerical Simulations of Molecular Sublimation}\label{sec:sublimation}
\begin{landscape}
\begin{figure}[htp]
\begin{center}
\includegraphics[width=\columnwidth]{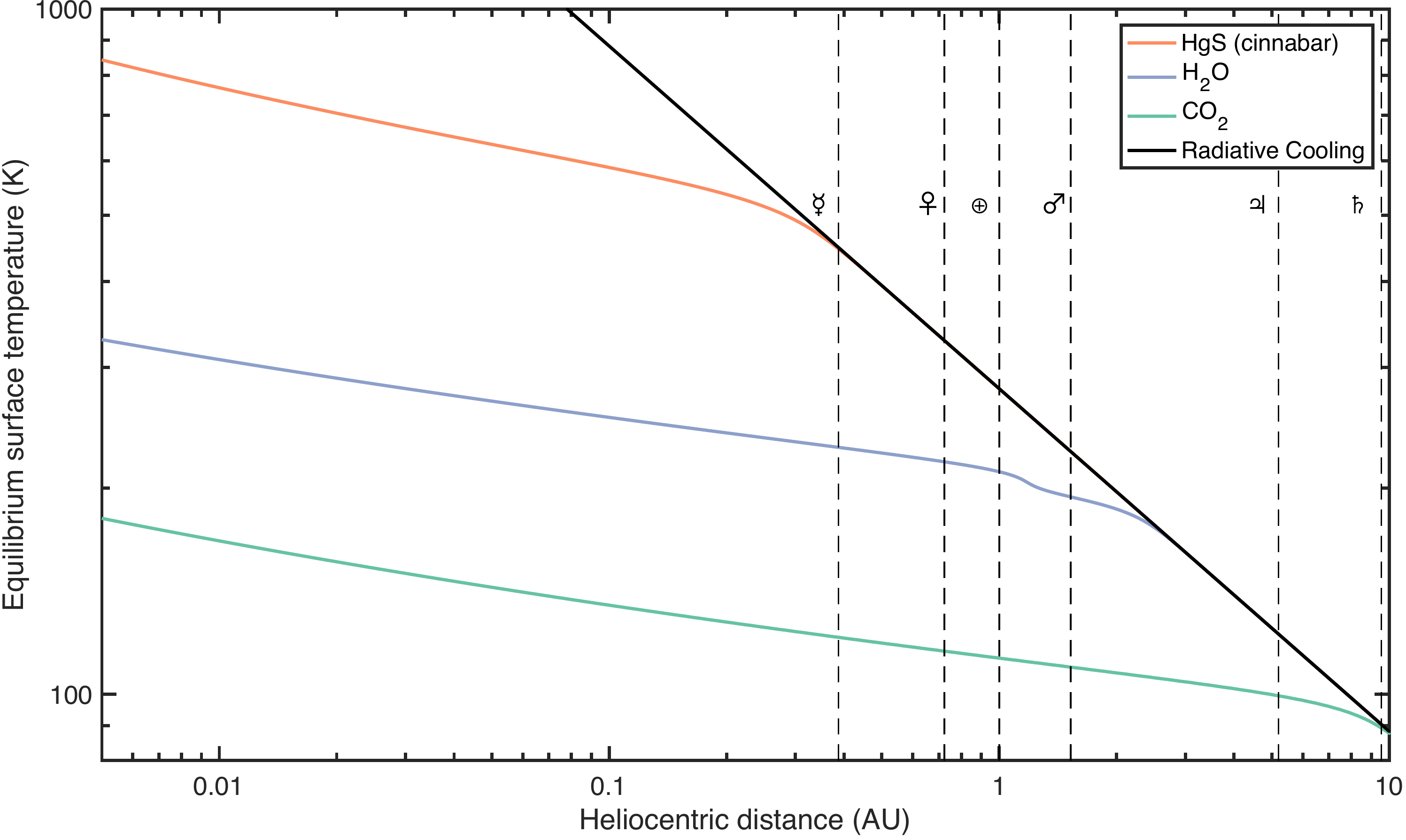}
\caption{Equilibrium temperature of materials versus heliocentric distance. Cinnabar (HgS) sublimates at high temperatures over a narrow distance range from the Sun; volatile materials (H$_2$O and CO$_2$) have a larger distance range where sublimation occurs. The diagonal black line shows the boundary below which radiative cooling dominates. The semi-major axes of the four inner and two gas giant planets are marked with vertical dashed lines.}
\label{fig:eqtempdistance}
\end{center}
\end{figure}
\end{landscape}

We numerically simulated the sublimation of cinnabar (HgS) to estimate the heliocentric distances at which various materials begin to sublimate and leave the surface of an asteroid, to better understand how labile elements would leave small, airless bodies in orbit around the Sun, in particularly Bennu and Phaethon.  
We used a numerical sublimation model \citep{Steckloff2015}, solving the energy balance equation at the surface of an object to compute the sublimation mass loss of a given material as a function of that material's physical properties (latent heat of sublimation, molecular mass, sublimation coefficient, and a laboratory-measured temperature and vapor pressure measurement).   
This model assumes a heterogeneous mixture of materials or pure material, such that the properties of one substance control the sublimation rate of that material from the surface of the asteroid. We incorporated sublimation of cinnabar as an example host phase of two labile elements, as well as H$_2$O and CO$_2$ (Figure \ref{fig:eqtempdistance}).  At large heliocentric distances, the surface cools predominantly through thermal radiation, leading to a steep slope in the sublimated mass loss curve. At smaller heliocentric distances the surface transitions to predominantly sublimative cooling, effective at dissipating heat,  causing the mass loss to flatten. The inflection point in each curve represents a heliocentric distance where sublimation becomes the dominant cooling mechanism.   


\section{Implications}

Quantifying the effects of thermal metamorphism on labile element mobilization in primitive chondritic meteorites has multiple implications.  Our results are essential for maintaining the pristine nature of samples returned from asteroid (101955) Bennu, constraining the behavior of labile elements in rock comets such as (3200) Phaethon, and extracting water from asteroids and lunar regolith for future in situ resource utilization.

\subsection{OSIRIS-REx Sample Return Mission to Asteroid (101955) Bennu}
\label{sec:orex}

The OSIRIS-REx (Origins, Spectral Interpretation, Resource Identification, Security-Regolith Explorer) asteroid sample return mission to asteroid (101955) Bennu (launched September 8, 2016) will return at least 60 g of regolith samples in 2023 \citep{Lauretta2017}. Bennu and (3200) Phaethon \citep[and Phaethon's likely parent body (2) Pallas; ][, see Section \ref{sec:phaethon}]{de2010}, show spectra similar to those of B-class asteroids \citep{Licandro2007,Clark2010,Clark2011}.  Several studies indicate the composition of Bennu is likely similar to that of carbonaceous chondrite meteorites and has remained relatively thermally unprocessed \citep{Clark2011,Muller2012}.  Laboratory studies of returned samples will allow for direct measurement of labile element depletion in material from what is considered a primitive body. Quantifying the loss of labile elements from Bennu over its dynamical history can inform models of the thermal effects of Bennu's orbital history.

Understanding the thermal response of elements in meteorites is essential for preserving the pristine nature of material sampled by OSIRIS-REx.  
We focus on a temperature range of 300 to 500 K, covering typical temperatures on Bennu \citep[300 to 400 K; ][]{Emery2014}, as well as what the OSIRIS-REx sample return container could experience during touch-and-go sampling and the spacecraft's return to Earth. We selected Murchison (type CM2), Nogoya (CM2), and Orgueil (CI1) for their similarity to the expected regolith composition of Bennu; we anticipate the mobilization of labile elements in these samples will provide information applicable to the returned sample from Bennu. 

\begin{landscape}
\begin{figure}[tb]

\begin{subfigure}[t]{0.5\hsize}
\centering    
\includegraphics[width=0.9\columnwidth]{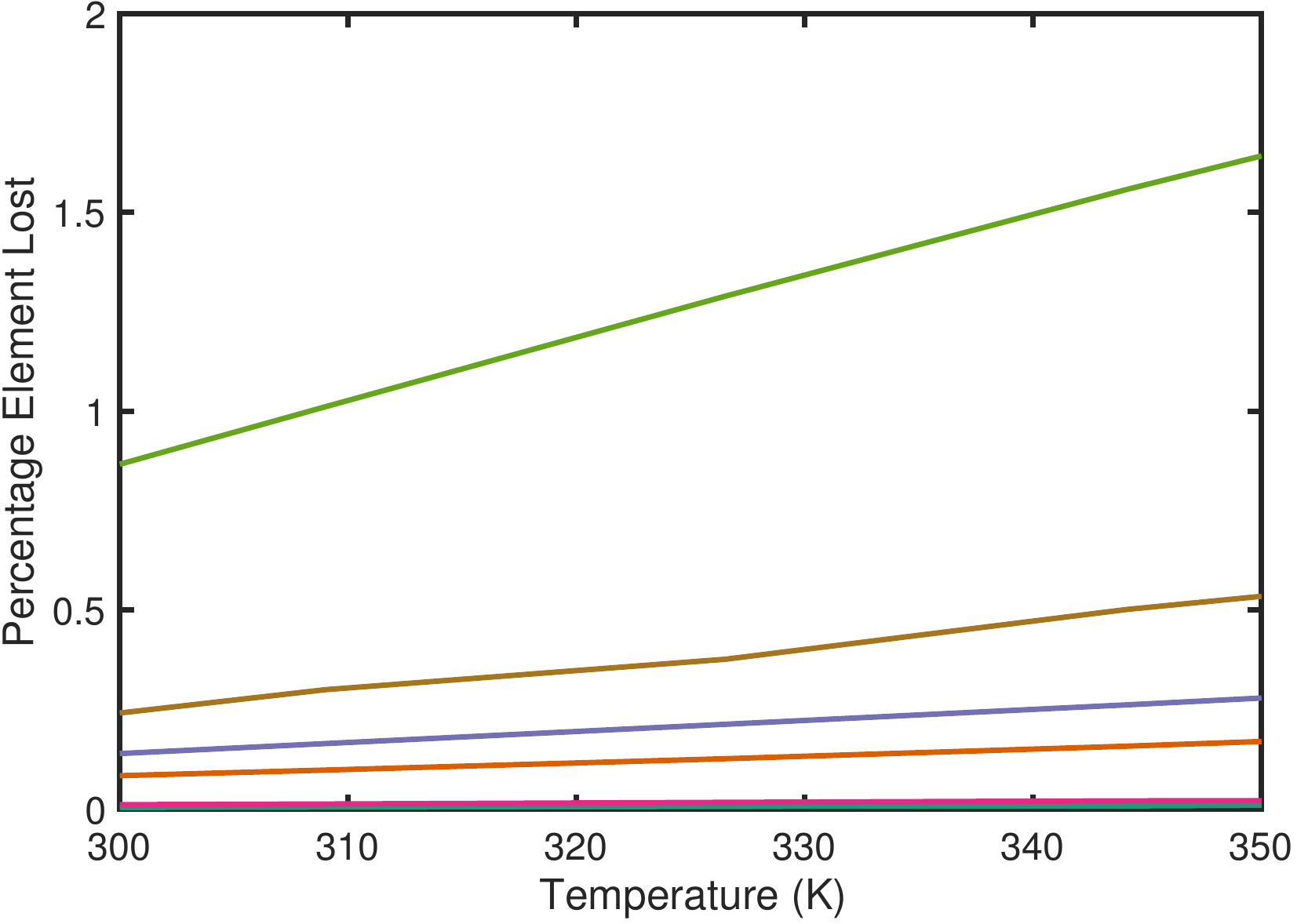}
    \caption{Orgueil}
\end{subfigure}   
\hfill
\begin{subfigure}[t]{0.5\hsize}
\centering    
\includegraphics[width=0.9\columnwidth]{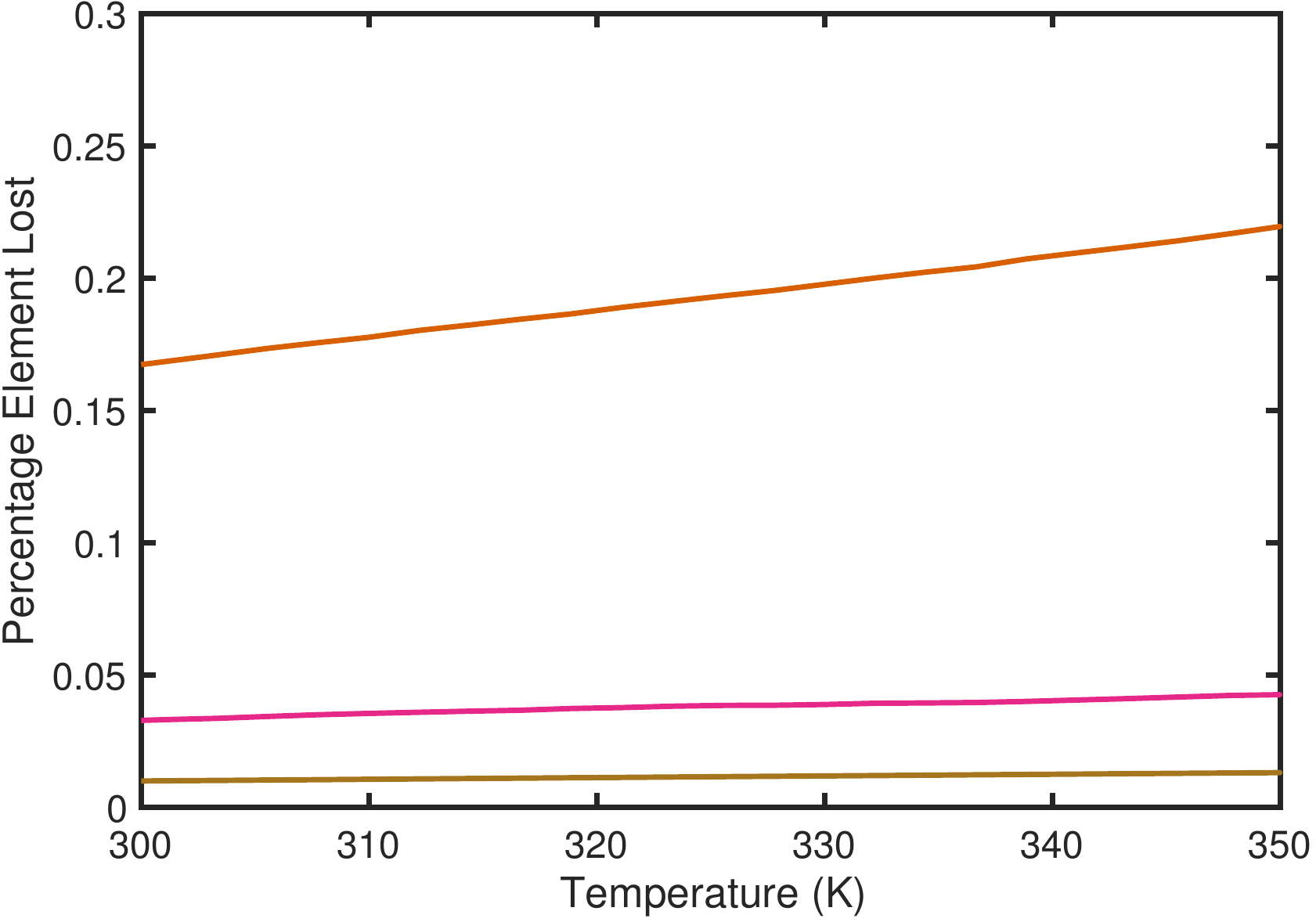}
    \caption{Murchison}
\end{subfigure}   
~
\begin{subfigure}[t]{0.5\hsize}
\centering      
\includegraphics[width=0.9\columnwidth]{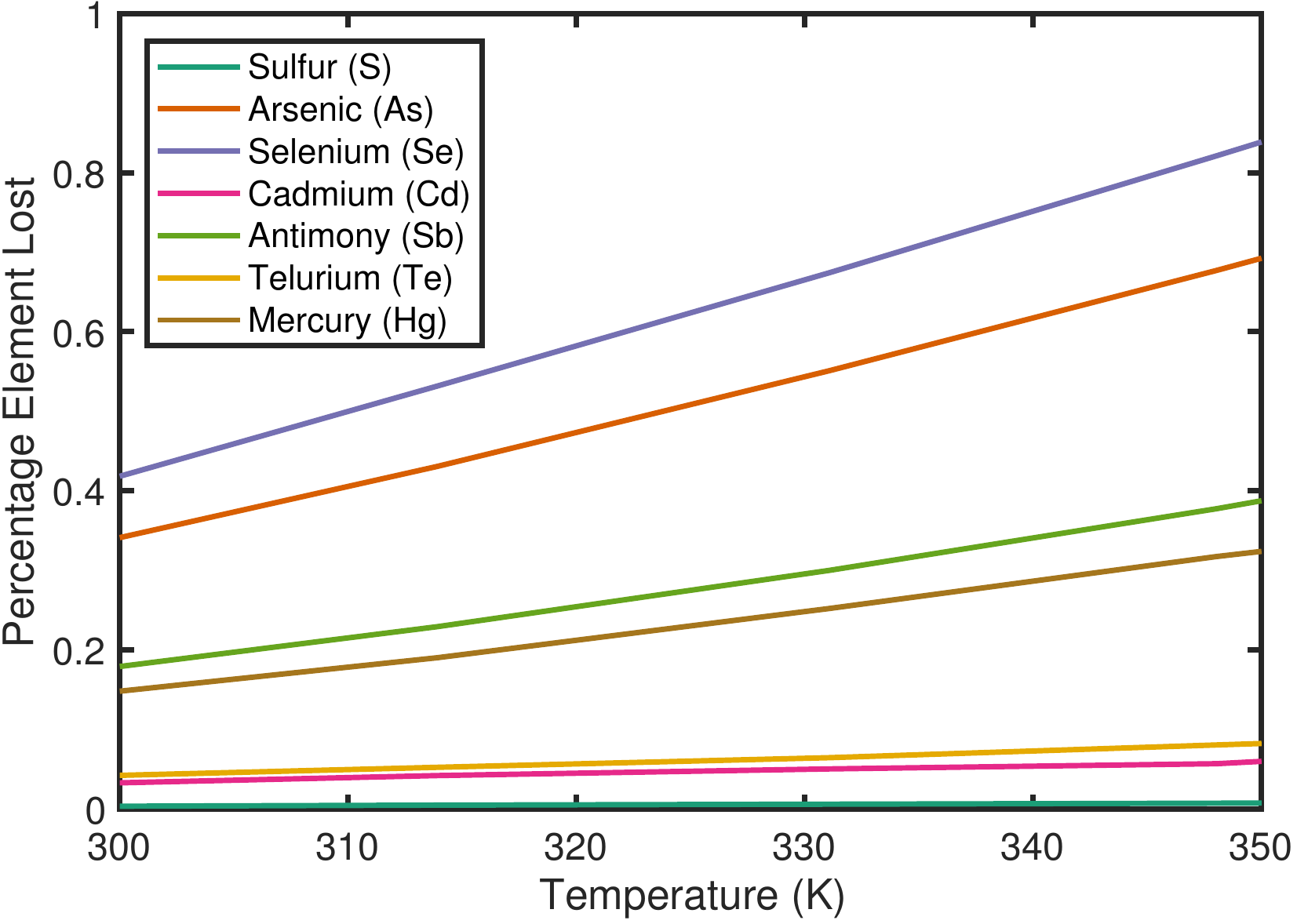}
	\caption{Nogoya}
\end{subfigure}   	
\hfill
\caption{Element percentages lost for Orgueil, Murchison, and Nogoya meteorite samples heated up to 350 K. Note the small percentages of elements lost over the 300--350 K range.  The vertical dashed line represents the maximum thermal limit (350 K) for the returned meteorite sample in the \textit{OSIRIS-REX} spacecraft sample return canister.  \citet{Emery2014} show a temperature range of 300--400 K for Bennu over a typical orbit.}
\label{fig:line350}
\end{figure}
\end{landscape}

\begin{landscape}
\begin{figure}[tb]

\begin{subfigure}[t]{0.5\hsize}
\centering    
\includegraphics[width=0.9\columnwidth]{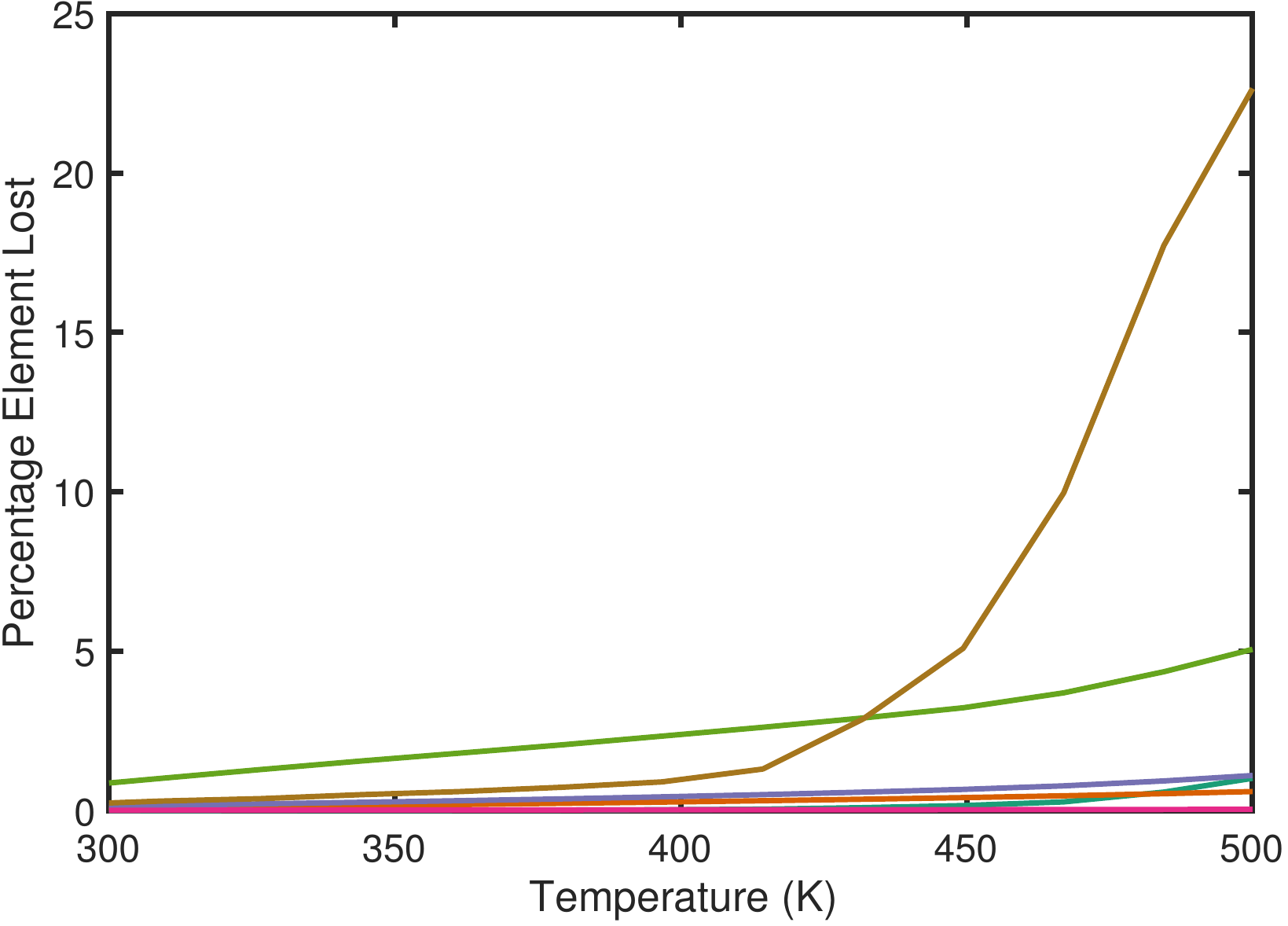}
    \caption{Orgueil}
\end{subfigure}   
\hfill
\begin{subfigure}[t]{0.5\hsize}
\centering    
\includegraphics[width=0.9\columnwidth]{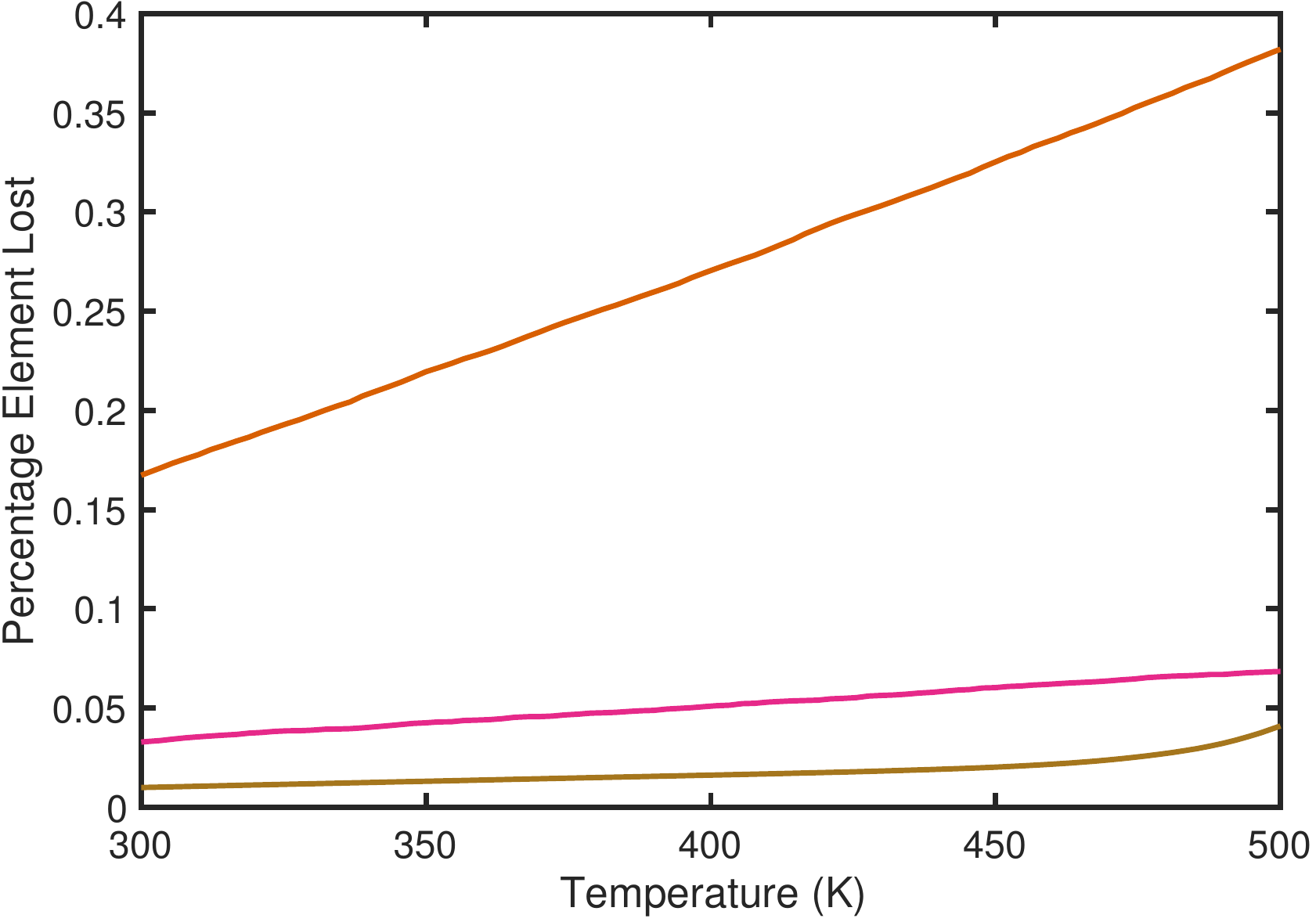}
    \caption{Murchison}
\end{subfigure}   
~
\begin{subfigure}[t]{0.5\hsize}
\centering      
\includegraphics[width=0.9\columnwidth]{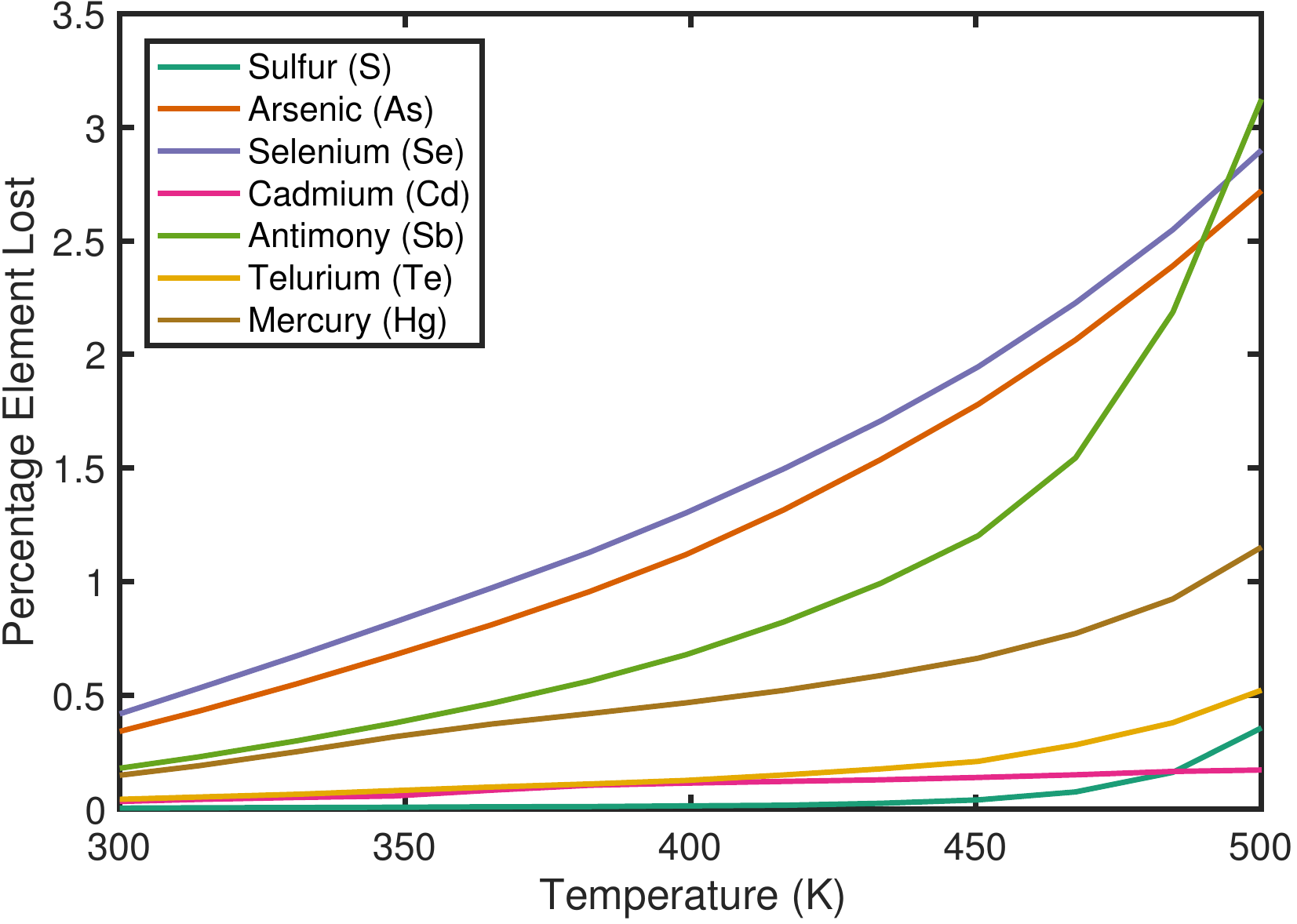}
	\caption{Nogoya}
\end{subfigure}  
\hfill
\caption{Element percentages lost for Orgueil, Murchison, and Nogoya meteorite samples heated up to 500 K. For the measured elements lost from Murchison and Nogoya, we see maximum losses over this temperature range from fractions of a percent to less than 5\%.  However, Orgueil shows loss of almost 23\% of S when heated to 500 K, unlike Nogoya.}
\label{fig:line500}
\end{figure}
\end{landscape}

Below 350 K, we observed a loss of up to 1.6\% of labile elements during heating of meteorite samples. Minimal S release was detected in this temperature range. We calculated the expected percentage losses for Murchison, Nogoya, and Orgueil meteorite samples up to 350 and 500 K (Table \ref{tab:element-loss}, Figures \ref{fig:line350} and \ref{fig:line500}). 

%

Should the sample experience temperature excursions of up to 500 K, we would expect to see considerable labile element loss, especially of Hg if the regolith has a similar composition to Orgueil.  Heating the Bennu sample past 400 K would irreversibly cause loss of elements of cosmochemical interest. To preserve pristinesamples from Bennu, keeping material below 350 K is a conservative, robust guard against the thermal release of labile elements. Maintaining the sample at a low temperature preserves labile elements in their original mineral phases and thus the thermally unaltered nature of these samples.

\citet{Delbo2011} calculated the probabilities of portions of Bennu's surface (including at the subsolar point, 50\% of the surface, and 50\% of the subsurface to a depth of 3 to 5 cm) being heated to specific temperatures. Combining their results with our heating experiments, we can predict the relative depletion of labile elements in the sample OSIRIS-REx returns from Bennu. If Bennu had a composition similar to that of Nogoya and was heated past 625 K, we would expect a loss of 15\% of Hg (Figure \ref{fig:delbo_murchison}). The \citet{Delbo2011} calculations correspond to past orbital histories, including previous perihelion distances of Bennu. The results of these laboratory heating experiments, when combined with future laboratory measurements of abundances and depletions of labile elements in the returned sample, will help constrain the past orbital behavior of Bennu. 

Hayabusa2 is a complimentary asteroid sample return mission to a carbonaceous asteroid, operated by the Japanese Aerospace Exploration Agency \citep{Watanabe2017}. This mission, which is presently exploring carbonaceous near-Earth asteroid (162173) Ryugu, will return $\sim100$ mg of sample to Earth in 2020, a sufficient quantity of material for analysis of mineralogy, organic chemistry, spectroscopy, space weathering processes, and major elements \citep{Sawada2017}.  Past ground-based spectral observations of Ryugu matched best with thermally altered CM chondrite Murchison, and CI chondrite Yamato 86029 \citep{Moskovitz2013}, both of which experienced aqueous alteration on their parent bodies.  

A small carry-on impactor on Hayabusa2 will create ejecta of subsurface materials on Ryugu, then the spacecraft will sample the resulting material, which, depending on the rate of impact gardening on Ryugu, would have undergone minimal heating due to sunlight exposure, or space weather \citep{Tachibana2014}.  Comparing the returned subsurface Ryugu samples with heated chondrites depleted in labile elements, such as those in this work, could constrain temperatures this material had been exposed to in the main asteroid belt, and possibly verify Ryugu's proposed origin from a background population of low-albedo main belt asteroids \citep{Campins2013}.

\subsection{Labile Element Sublimation from Rock Comets such as (3200) Phaethon} \label{sec:phaethon}
One potential laboratory for studying labile element loss with temperature is asteroid (3200) Phaethon, associated with the Geminid meteor shower.  Phaethon, an object with a perihelion distance of 0.14 and a high eccentricity (e = 0.89), Phaethon experiences temperatures of 800 to 1100 K \citep{Ohtsuka2009}; thus, thermal effects on this body should contribute to material loss--whether the progenitor meteoroids of the Geminid shower, or labile elements.  

An active Phaethon could resupply the Geminid stream, and observers documented post-perihelion tail formation and brightening of Phaethon in 2009 and 2012 \citep{Jewitt2010,Li2013,Jewitt2013}. Several brightening and tail-forming mechanisms are possible; ultimately, these events were attributed to dust originating from the surface of the asteroid caused by thermal fracture and cracking owing desiccation of hydrated minerals on the surface. However, the dust size and mass calculated to be ejected from the surface of Phaethon are insufficient to resupply Geminids meteor stream \citep{Jewitt2013,Ye2018}.  The Geminids meteoroid stream itself is dynamically younger than Phaethon, on the order of $10^3$ years \citep{Ohtsuka2009}, meaning the dust in the Geminid stream must have been produced recently from the surface of Phaethon.  

Phaethon is 5.1 km in diameter \citep{Tedesco2004}.  Lightcurve observations indicated a rotation period of 3.6 hours \citep{Pravec1998} and a nearly spherical shape with some irregularities \citep{Ansdell2014,Hanus2016,Kim2018}. Analysis of radar data from the Arecibo Observatory Planetary Radar System shows Phaethon to be roughly spherical and $\sim6$ km in diameter, with some topographic irregularities \citep{Taylor2018}.  \citet{Ansdell2014} modeled a highly oblique pole orientation for Phaethon; thus, volatiles could be activated during perihelion as the north pole, shadowed during most of Phaethon's orbit, is suddenly oriented toward the Sun. Polarimetry studies of Phaethon's surface suggest variation in the thickness of surface regolith, large grain sizes in the regolith, surface porosity, or topography \citep{Borisov2018,Ito2018}.  Sublimation combined with thermal torques affecting spin axis \citep{Steckloff2016} may cause pole realignment, in additional to the effects of ``traditional'' YORP (Yarkovsky-O'Keefe-Radzievskii-Paddack) torques \citep{Vokrouhlicky2002}, allowing for cooler core temperatures as well as retention of pristine subsurface volatiles.

Mass loss from Phaethon is not likely due to ice sublimation \citep{Jewitt2013}; other mass loss mechanisms must be at play, given that surface ice would be thermodynamically unstable on Phaethon because of the high surface temperature \citep{Jewitt2018}.  Ice at the core of Phaethon would be ``phase-lagged from the surface heat, leaving no explanation for the coincidence between activity and perihelion'' \citep{Jewitt2010}.  Thermal fracture and fatigue could be responsible for cracking the rocky material on the surface of Phaethon and producing dust \citep{Jewitt2013}; indeed, $\sim100$-MPa stresses can be induced in microstructures on airless bodies similar to Phaethon \citep{Molaro2015}, and thermal cycling could cause crack propagation, producing rocky material, perhaps comparable in total mass to the Geminid meteor stream.  

Phaethon is a B-class asteroid based on near-infrared spectra, and this object likely has a composition similar to those of aqueously altered primitive meteorites such as CI or CM chondrites \citep{Licandro2007,Kinoshita2017}. Other workers have established a correlation between B-class asteroid spectra and experimentally thermally metamorphized Murchison meteorite samples; \citet[][ and references therein]{Clark2010} show that a CK4 chondrite may be the closest match to Phaethon.  \citet{de2010} established a ``compositional and dynamical connection'' between Phaethon and main-belt asteroid (2) Pallas, also a B-class asteroid; recent observations support this connection \citep{Devogele2018}. Additionally, \citet{Clark2011} showed spectral similarities between Phaethon and B-class asteroids such as Bennu and Pallas; these findings are underscored by those of \citet{Takir2018}. 

We showed Hg is a potential tracer of thermal metamorphic history in primitive meteorites; rock comets composed of material similar to primitive meteorites could be heated to hundreds of Kelvin based on temperatures reached by small objects with perihelia of Mercury-crossing small bodies, or sungrazer comets \citep[][see Figure \ref{fig:eqtempdistance}]{Knight2010}.  Hg is the first element released from the meteorites during heating making it the most labile metal in carbonaceous chondrites, and it would sublimate at temperatures below the equilibrium temperature for a body with Phaethon's perihelion of 0.14 au if, for example, it was released from cinnabar (Section \ref{sec:sublimation}). 
In all meteorites we studied, 99\% of the total Hg released by 700 K. The majority of Hg released at low temperatures is likely present as adsorbed or intergranular Hg. Mercury release from the carbonaceous chondrites continued at low levels between 700 and 1200 K. A refractory material, such as a Fe-based alloy or ferromagnesian silicates, may host this Hg. The slow rate of Hg release may result from Hg diffusion out of these phases. \citet{Meier2016} studied the release of Hg from a variety of meteorite materials, including carbonaceous chondrites. Their work shows a significant release of Hg below 425 K, which they interpret as the phase transition between troilite to pyrrhotite. 

In our experiments, most Hg loss occurs below 650 K (except for Nogoya, in which the release of Hg is more continuous than peaked release of Hg above 450 K, and Murray, where the total amount of Hg lost relative to an initial amount was not measured). All studied meteorites, except for Murray (Figure \ref{fig:abundances}), contain a substantial low-temperature component released below 700 K. Observations of the LCROSS lunar impact plume showed abundant Hg in a body as evolved as the Moon \citep{Gladstone2010}; it may be possible to detect Hg on less processed bodies such as rock comets. Future space telescopes equipped with UV spectrometer capabilities could potentially detect strong UV Hg lines on rock comets such as Phaethon or other ``activated asteroids'' \citep{Hsieh2006}.

Our samples became rapidly depleted in labile elements and, in particular, lost $\sim75$\% of their Hg content when heated from $\sim500$ to 700 K, which corresponds to heliocentric distances of $\sim0.15--0.3$ au, consistent with our thermal models. For instance, Hg has strong emission lines in the UV (185 nm); thus its presence (or absence) relative to carbonaceous chondrite abundances would indicate whether these bodies had perihelia in their dynamical histories inside of 0.15 au and therefore may have previously been Phaethon-like rock comets.  Observations of the Geminid meteor shower in the UV (300 to 600 nm) show depletion of sodium relative to solar abundances, likely owing to thermal alteration of alkali silicates \citep{Kasuga2005}.  On asteroids with perihelia less than that of Phaethon, alkali metals may be depleted, as well as labile elements; indeed, more Na depletion is expected for meteor showers with perihelia of less than 0.1 au \citep{Kasuga2006b}.
Phaethon will continue to be a target of remote observing via future space telescopes, or perhaps balloon-borne observing platforms.  Additionally, several \textit{in situ} missions have been proposed to Phaethon to study its surface and connection with the Geminids \citep{Kasuga2006,Quarta2010,Sarli2015,Arai2018}.  Equipping such a mission with a UV spectrometer could potentially detect the presence or absence of strong UV Hg lines on rock comets or rock comet candidates.

\subsection{\textit{In Situ} Resource Utilization}
\label{sec:isru}
For future in situ resource utilization, knowing the abundance of volatile elements in a primitive meteorite, along with the percentage of water, would be useful for water purification processes to create both drinking water and rocket propellant. Heating primitive meteorites such as Murchison to release water vapor will also release high concentrations of labile elements, many of which exceed Environmental Protection Agency limits for contaminants in drinking water \citep[][, Table \ref{tab:h2oconcentration}]{Kelsey2013}. Water mined from extraterrestrial sources has long been predicted to be polluted with labile elements such as Hg \citep{Reed1999}; however, this could be mitigated by adsorption via ion exchange resins \citep{Chiarle2000}.  Producing water with high concentrations of Hg to make hydrogen gas for fuel could cause metal failure from brittle fracture in the hydrolyzer system, considering that Hg can cause brittleness in aluminum in natural gas processing plants \citep{Nichols1961}.

Further, high concentrations of sulfur in water vapor can poison platinum electrodes in traditional electrolyzers used to purify water vapor, leading to irreversible losses in platinum electrode performance \citep{Kelsey2013}. The deactivation of Pt sites is independent of the type of sulfur contamination \citep{Sethuraman2010}. Purification systems using substances such as Nafion should be theoretically impermeable to sulfur compounds resulting from vaporization of primitive meteorite material \citep{Yeo1980,Kelsey2013}.  For future in situ resource utilization, future work is needed to enable purification of water from heavy metals and sulfur without damaging the purification apparatus.

\section{Conclusions}
We find Hg is the most volatile metal in carbonaceous and ordinary chondrites we studied. Of the other volatile siderophile or chalcophile trace elements investigated, only As, Se, Cd, Sb, and Te released below 775 K. In many cases, the release of these elements correlates with that of S. Thermal metamorphism at low temperatures could produce fractionation of thermally labile elements. The fractionation would not follow a simple volatility trend; instead, it would reflect both the different mobilities of these elements during aqueous processes and their volatility during thermal metamorphism. It is likely that the release of these elements also varies with atmospheric conditions, such as atmospheric composition or oxygen fugacity, for example.  

In almost all the meteorites that we analyzed, we found evidence for the presence of a low-temperature phase that serves as a host site for As, Se, Sb, Te, and (rarely) Cd.  The correlation with the release of S indicates that this phase is a sulfide or sulfate. The presence of this phase may be an excellent indicator that low-temperature processes occurred on meteorite parent bodies.

This work focused on a suite of seven labile elements, but laboratory measurements exist for eight more, including Zn, In, Sn, Pt, Au, Tl, Pb, and Bi. Future analysis could extend our techniques to these additional elements, providing more points of comparison to subsequent work investigating the thermal behavior of these elements in meteorites and parent-body material.  
Using x-ray diffraction analysis and/or microscopic observations to compare samples quenched at a variety of temperatures during heating with unheated samples of the same meteorite could help quantify heating processes and reactions.  Examining in broader terms how chondrite material heats would provide a deeper understanding of how thermal processing affects asteroid surfaces, including phase changes and dehydration.  

Connecting the temperature release results with known meteorite mineral phases \citep{Rubin1997} would further our understanding of how material on meteorites thermally evolves through various phases.  This work is important for understanding the temperature constraints for future sample return missions, as well as tracing the heating history of objects as they move through the Solar System.  

Going forward, detecting the signature (or absence thereof) of labile elements such as Hg via remote sensing techniques would allow the connection of space-based measurements with laboratory experiments to constrain the orbital evolution of small Solar System bodies.

\section*{Acknowledgements}
This work was supported by NASA grants NAG5-11355 (DSL) and NC01-109 (JDB) and by NASA contract NNM10AA11C issued through the New Frontiers Program (AS and DSL).  Carleton Moore and the Center for Meteorite Studies at ASU provided the samples.  Bulk analysis was conducted at the University of Arizona; thermal release experiments were performed at the University of Michigan.  AS thanks Ryan B.~Anderson, Julia Kamenetzky, James Tuttle Keane, Daniel Y.~Lo, Zachary A.~LaBry, Margaret E.~Landis, Moses P.~Milazzo, Henry Ngo, Alex H.~Parker, Erin L.~Ryan, Molly N.~Simon, Maria E.~Steinrueck, Nick Wirtz, Catherine Wolner, and Brian C.~Wolven for helpful comments on this manuscript, as well as data analysis and figure presentation.  
Plot colors are from \href{http://colorbrewer.org/}{ColorBrewer.org}, funded by the NSF Digital Government program during 2001--02 and designed at the GeoVISTA Center at Pennsylvania State University by Cynthia A.~Brewer (National Science Foundation grants 9983451, 9983459, and 9983461).  
We made use of NASA's Astrophysics Data System Bibliographic Services as well as the AstroBetter blog and wiki.  
We thank the anonymous reviewers for their thorough reading of and constructive feedback on our manuscript.  

\section*{References}
\bibliography{biblio}

\clearpage
\section{Tables}
\begin{landscape}
\centering
\begin{table}[]
\caption{Experimental studies of volatile-element loss during thermal metamorphism.}
\label{tab:metamorphism}

\begin{tabular}{p{1in} | p{0.8in} p{0.7in} p{0.7in} p{2in} p{2in}}
Meteorite                & Temperature (K)      & Atmosphere    & Elements                                       & Results                                                                                                                                                                            & Reference                                \\
\hline
Allende (CV)             & 700 to 1300 K & H$_2$            &                                                & Co and Ga retained. At 1200 K loss of $\sim$70\% In, 80\% Bi, 92\% Tl, \textless{}10\% of Se.                                                                                       & \citet{Ikramuddin1975gca}            \\
Allende (CV)             & 700 to 1300 K & H$_2$, He, and O$_2$ &                                                & \textgreater{}1000 K signicant loss of pentlandite and troilite.                                                                                                                    & \citet{Ikramuddin1975nat}                   \\
Murchison (CM)           &               &               & Co, Ga, Se, Bi, Tl, In, Ag, Cs, Te, Zn, and Cd & At 1200 K \textgreater{}90\% of Bi, Tl, In, Te, Ag, Cd, and Zn and 25\% of Ga and Se released.  Cd, Ga, In, and Se lost from single site;  Ag, Bi, Te, Tl, and Zn lost from two sites. & \citet{Matza1977volatile}                 \\
Allende (CV)             & up to 1700 K  &               & adding As, Cu and Sb                           & All three elements retained at low Ts. Incipient loss of As, Cu, and Sb \textgreater{}1200 K.                                                                                        & \citet{Bart1980, Ngo1980} \\
Ten CI and CM chondrites &               & pure O$_2$       & S                                              & 7 different components release S. CIs contain more oxidized S than CMs.                                                                                                               & \citet{Burgess1991}                     
\end{tabular}
\end{table}
\end{landscape}

\begin{table}[]
\caption{Meteorite element percentage loss to up 350 and 500 K}
\begin{center}
\begin{tabular}{l | c c c c c c c}
\multirow{2}{*}{Meteorite} & \multicolumn{7}{c}{Element percentage lost up to 350 K} \\
                  & S           & As    & Se   & Cd  & Sb     & Te     & Hg   \\
\hline
Murchison &   -           & 0.22 &  -     & 0.04 &  -      &  -      & 0.01 \\
Nogoya     & $<$0.01 & 0.69 & 0.84 & 0.06 & 0.39 & 0.08 & 0.32 \\
Orgueil      & $<$0.01 & 0.17 & 0.28 & 0.02 & 1.62 & -       & 0.52 \\

\multicolumn{8}{c}{~}\\
\multirow{2}{*}{Meteorite} & \multicolumn{7}{c}{Element percentage lost up to 500 K}\\
                  & S           & As    & Se   & Cd  & Sb     & Te     & Hg   \\
\hline
Murchison & -       & 0.38 & -       & 0.07 & -       & -      & 0.04\\ 
Nogoya     & 0.33 & 2.71 & 2.88 & 0.17 & 3.06 & 0.51 & 1.12\\
Orgueil      & 0.97 & 0.59 & 1.09 & 0.04 & 4.99 & -      & 22.51\\

\end{tabular}
\label{tab:element-loss}
\end{center}
\end{table}

\begin{table}[]
\caption{Trace-element content of water released from the Murchison carbonaceous meteorite at 1000 K.  Sulfur, Hg, and Pb show high concentrations.  From \citet{Kelsey2013}, used with permission.}
\begin{center}
\begin{tabular}{p{1in} p{2in}}

Element & Concentration with respect to water release (mg/L) \\
~& ~\\
S  & 50000 \\
Zn & 15    \\
As & 0.41  \\
Se & 0.95  \\
Cd & 2.7   \\
In & 3.7   \\
Sn & 2.8   \\
Sb & 0.02  \\
Te & 0.11  \\
Au & 0.83  \\
Hg & 7.9   \\
Tl & 0.51  \\
Pb & 27    \\
Bi & 0.31  \\

\end{tabular}
\end{center}
\label{tab:h2oconcentration}
\end{table}

\clearpage
\section{Figures}
\newcommand{\AbunCaption}{Element abundances lost versus temperature for the 12 meteorite samples.  The data are scaled by the maximum number of counts for each element.}
\newcommand{\AbunWidth}{0.8}
\newcommand{\SubAbunWidth}{0.45}

\begin{landscape}
\begin{figure}[tb]
\begin{subfigure}[t]{\SubAbunWidth\hsize}
\centering    
\includegraphics[width=\AbunWidth\linewidth]{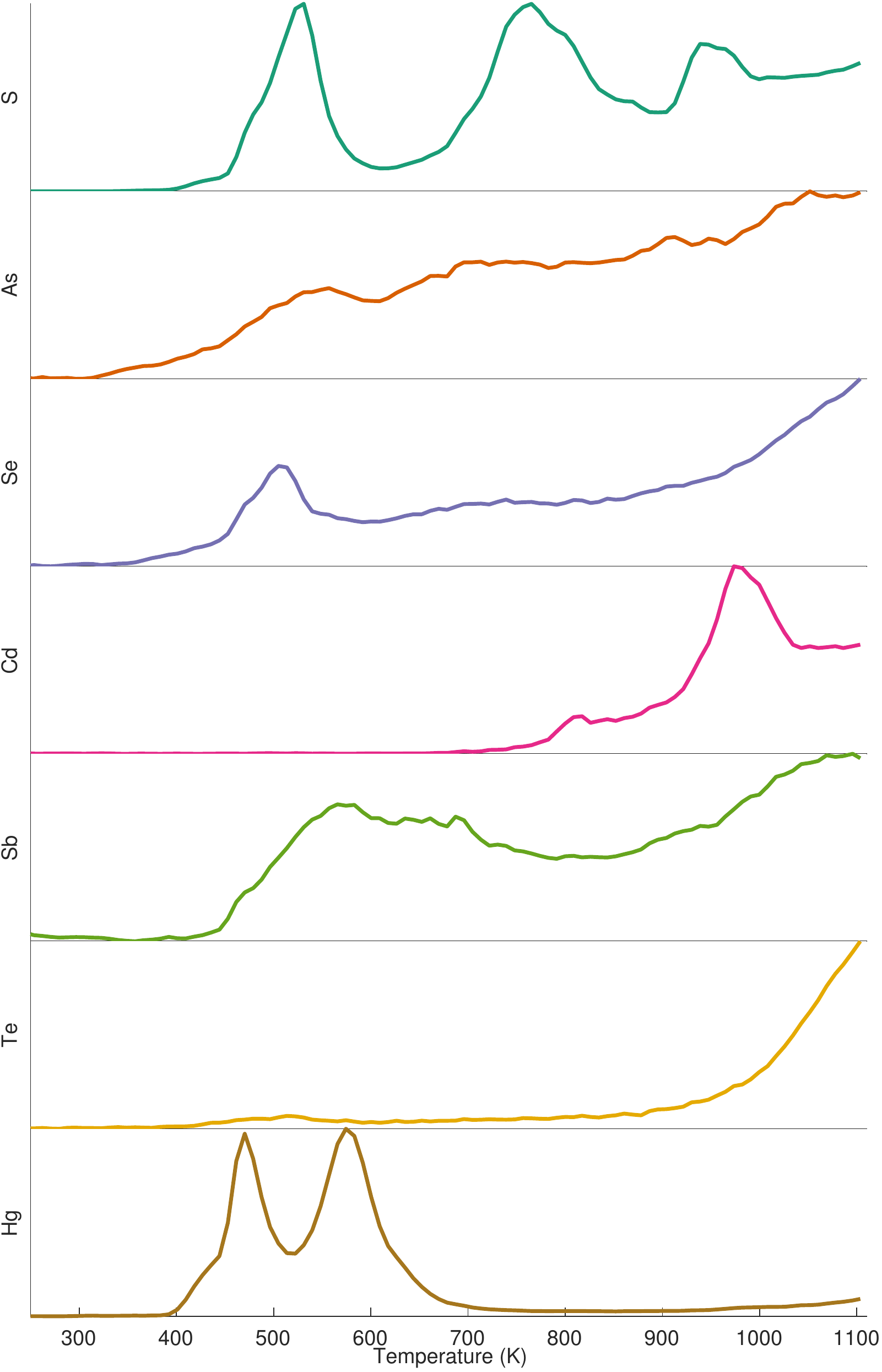}
    \caption{Orgueil: There are three distinct maxima in the S release profile at 530, 765, and 940 K.  The first peak falls between the two-peak release profile of Hg (470 and 575 K), perhaps owing to decomposition of troilite.  Hg release slowed between 750 and 900 K and increased slowly above 900 K.   Other element release profiles have peaks within 35 K of the first S peak (As, 550 K; Se, 510 K; Sb, 565 K).  Se released continuously above 835 K, Te above 650 K.}
    \label{fig:ab-orgueil}
\end{subfigure}   
\hfill
\begin{subfigure}[t]{\SubAbunWidth\hsize}
\centering    
\includegraphics[width=\AbunWidth\linewidth]{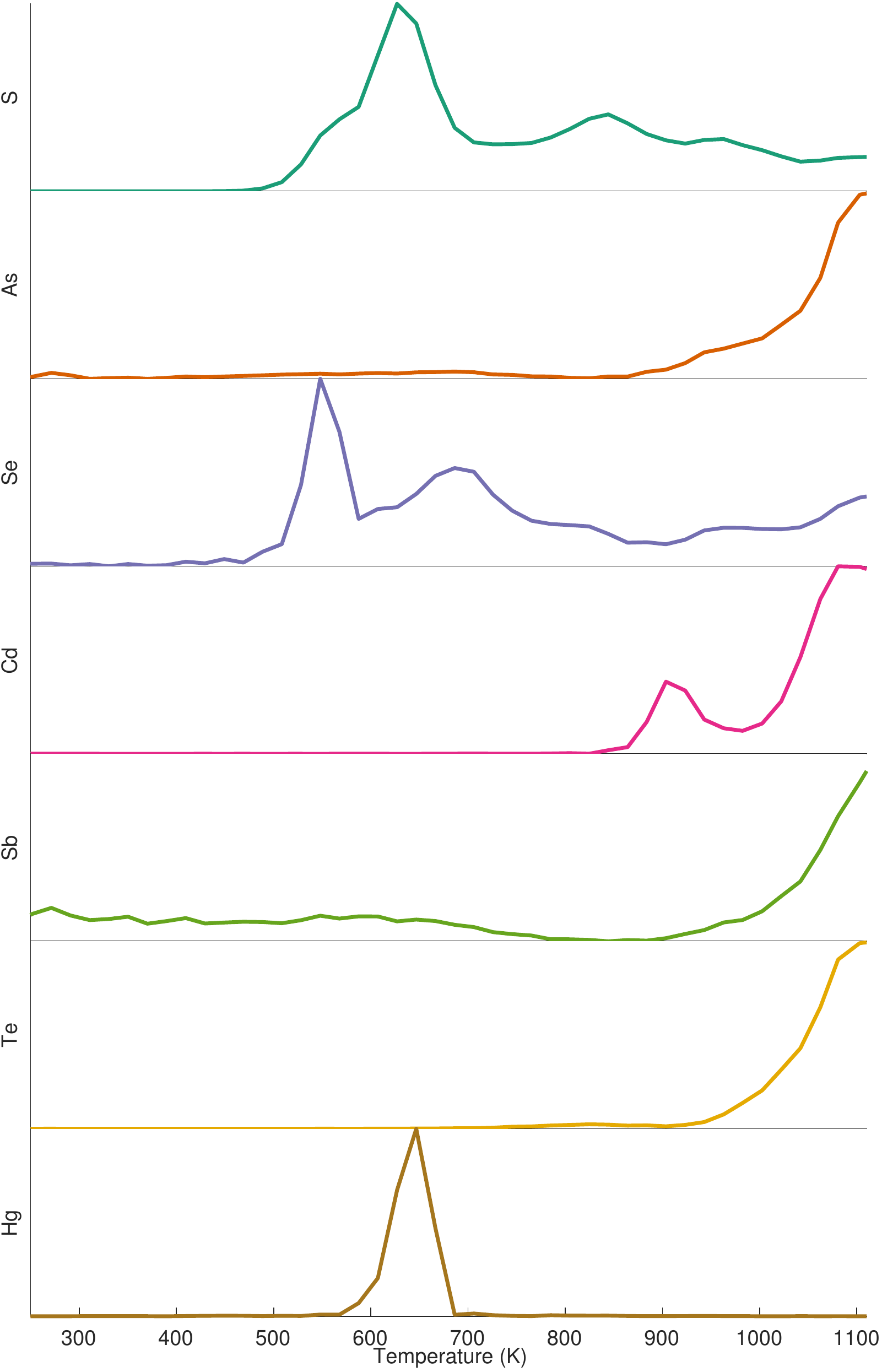}
    \caption{Murchison: S release peaks at 625 K, with smaller peaks at 845 and 960 K.  As release is initially flat before increasing after 865 K.  Se has a large peak at 545 K, and a smaller one at 890 K.  Cd and As show little release until 845 K, peaking at 900 and 1080 K.  Sb release decreases until increasing at 845 K.  Te, like As and Cd, increases release above 900 K.  Hg peaks at 645 K.}
        \label{fig:ab-murchison}
\end{subfigure}
\caption{\AbunCaption}
 \label{fig:abundances}
\end{figure}
\end{landscape}

\clearpage   
 
\begin{landscape} 
\begin{figure}[tb]\ContinuedFloat
\begin{subfigure}[t]{\SubAbunWidth\hsize}
\centering    
\includegraphics[width=\AbunWidth\linewidth]{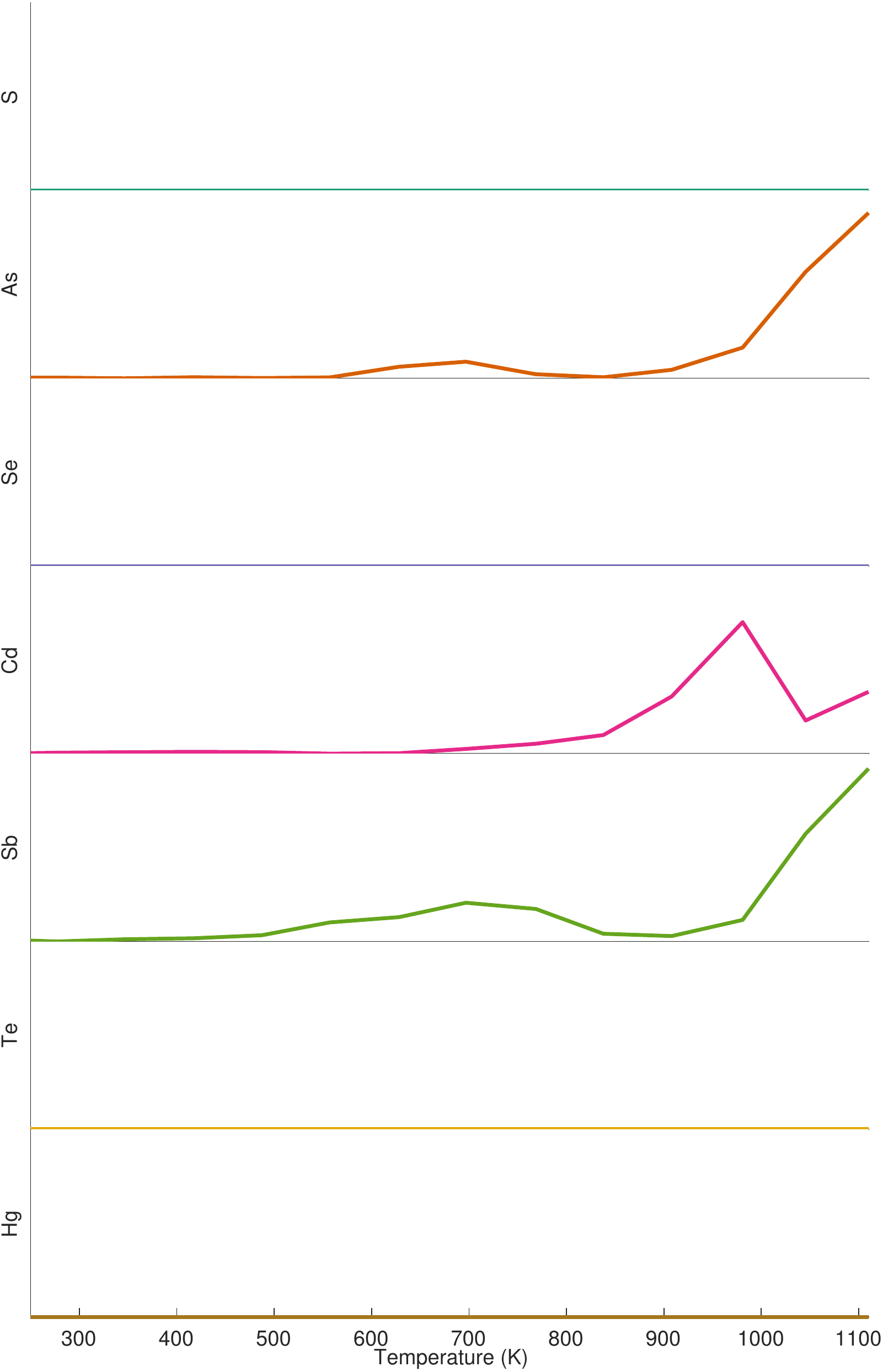}
    \caption{Murray: this experiment was run at a lower resolution than other samples.  We did not detect S, Se, Te, or Hg. As has small peak at 700 K and increases after 900 K. Cd has a peak at 980 K. Sb, similar to As, has a small peak at 700 K, then increases after 900 K.}
        \label{fig:ab-murray}
\end{subfigure}   
\hfill
\begin{subfigure}[t]{\SubAbunWidth\hsize}
\centering    
\includegraphics[width=\AbunWidth\linewidth]{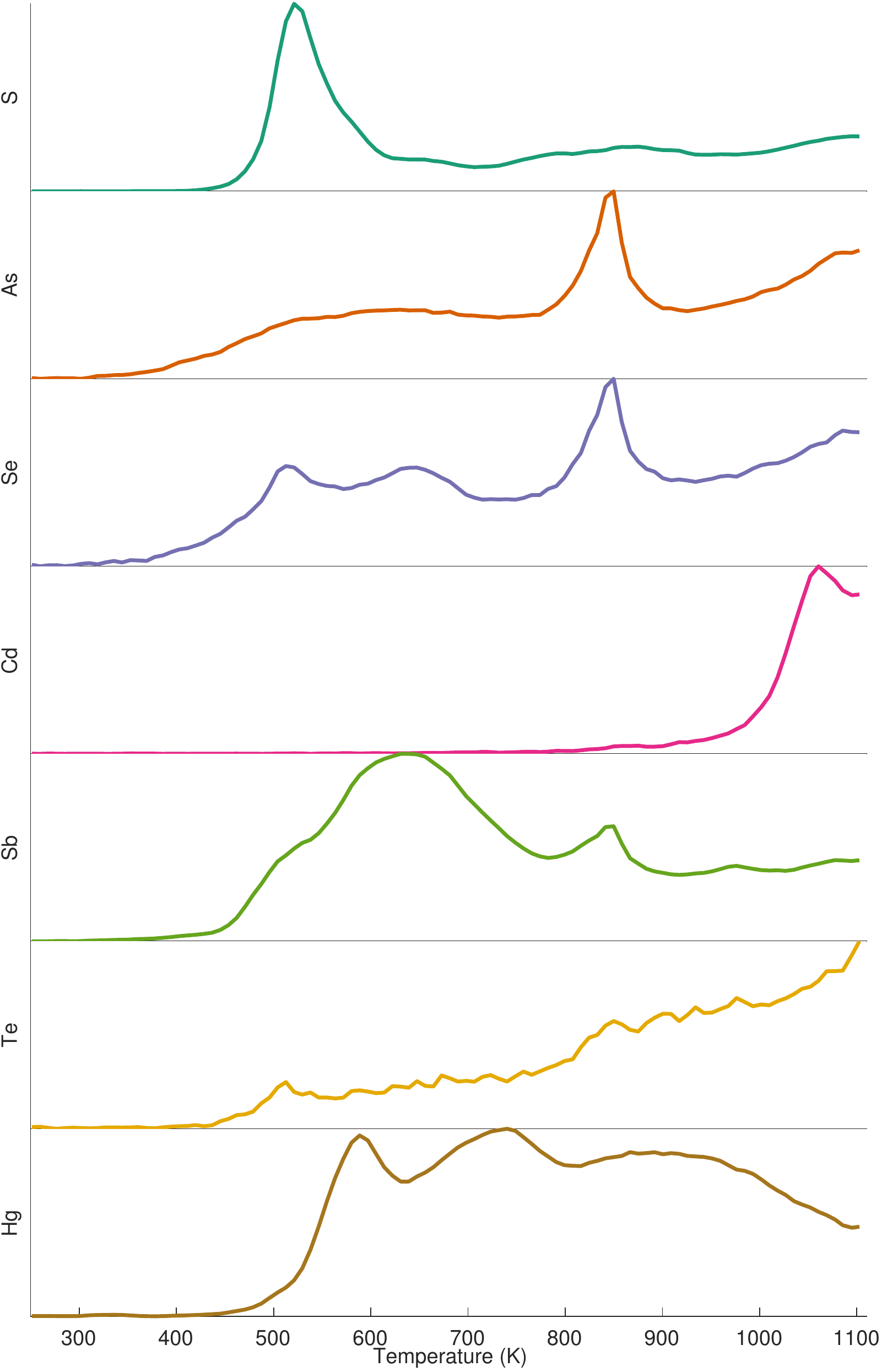}
    \caption{Nogoya: S has a prominent peak at 520 K. This corresponds to a medium peak in Se at 515 K, in Te at 510 K.  As has a sharp peak at 850 K, corresponding to peaks in Se and Sb.  Se has a peak at 640 K, corresponding to a large peak in Sb release.  Cd has a peak at 1060 K. Te releases continuously after 435 K. Hg has peaks at 590, 740, and 910 K.}
        \label{fig:ab-nogoya}
\end{subfigure}
    \caption{\AbunCaption}
    \end{figure}
\end{landscape}    

\clearpage   
 
\begin{landscape} 
\begin{figure}[tb]\ContinuedFloat
\begin{subfigure}[t]{\SubAbunWidth\hsize}
\centering    
\includegraphics[width=\AbunWidth\linewidth]{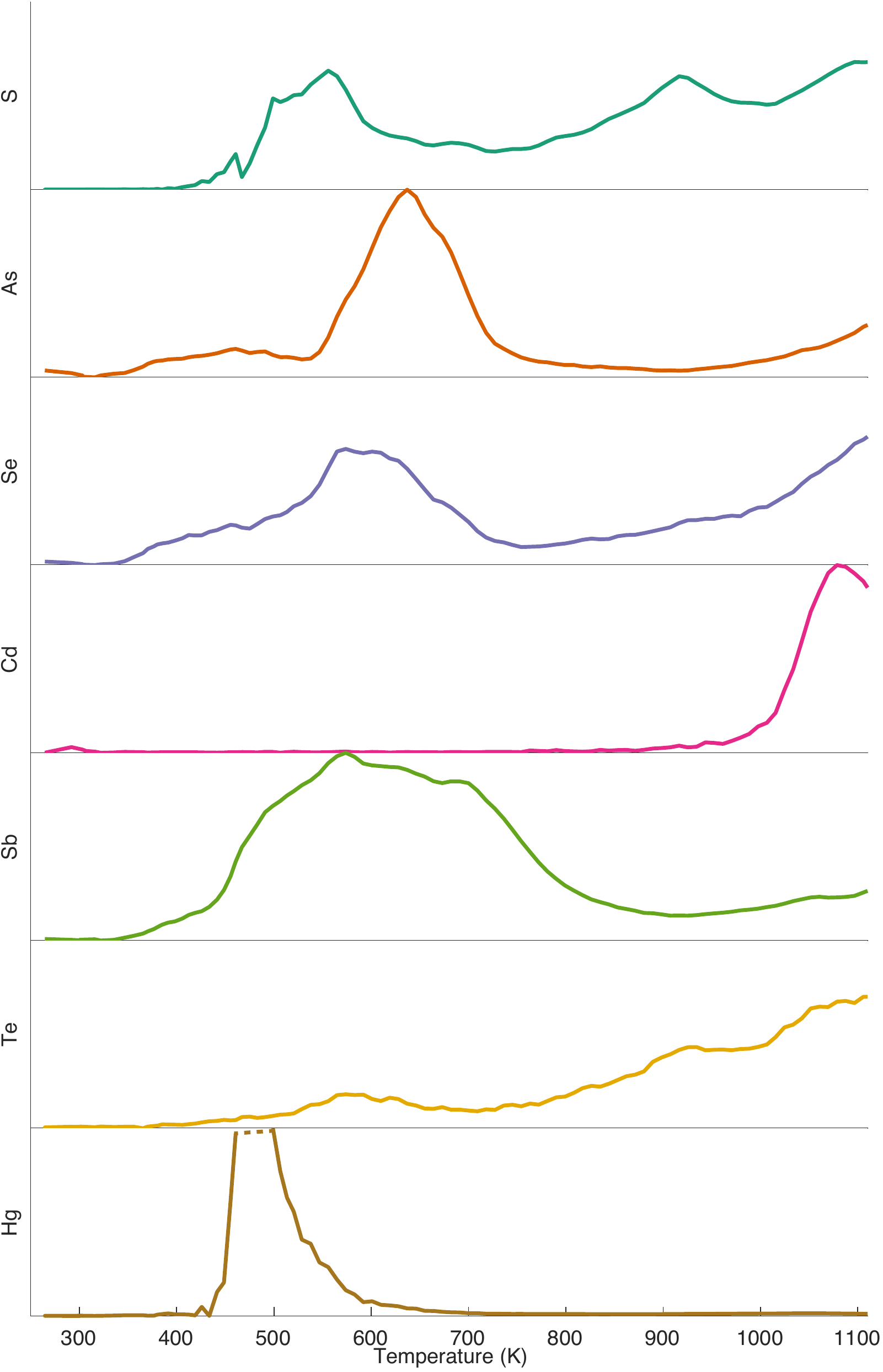}
    \caption{Cold Bokkeveld: S peaks at 560, 915, and 1100 K. As has a peak in release at 635 K, and Se at 575 K. Both As and Se increase in countse above 915 and 765 K, respectively. Cd has a mostly flat release profile and increases above 875 K, with a peak at 1080 K. Sb has a wide peak at 575 K. Te increases slowly throughout the observed temperature range. Hg peaks between 460 and 500 K; the dashed line represents detector saturation.}
    \label{fig:abuncoldbok}
\end{subfigure}   
\hfill
\begin{subfigure}[t]{\SubAbunWidth\hsize}
\centering    
\includegraphics[width=\AbunWidth\linewidth]{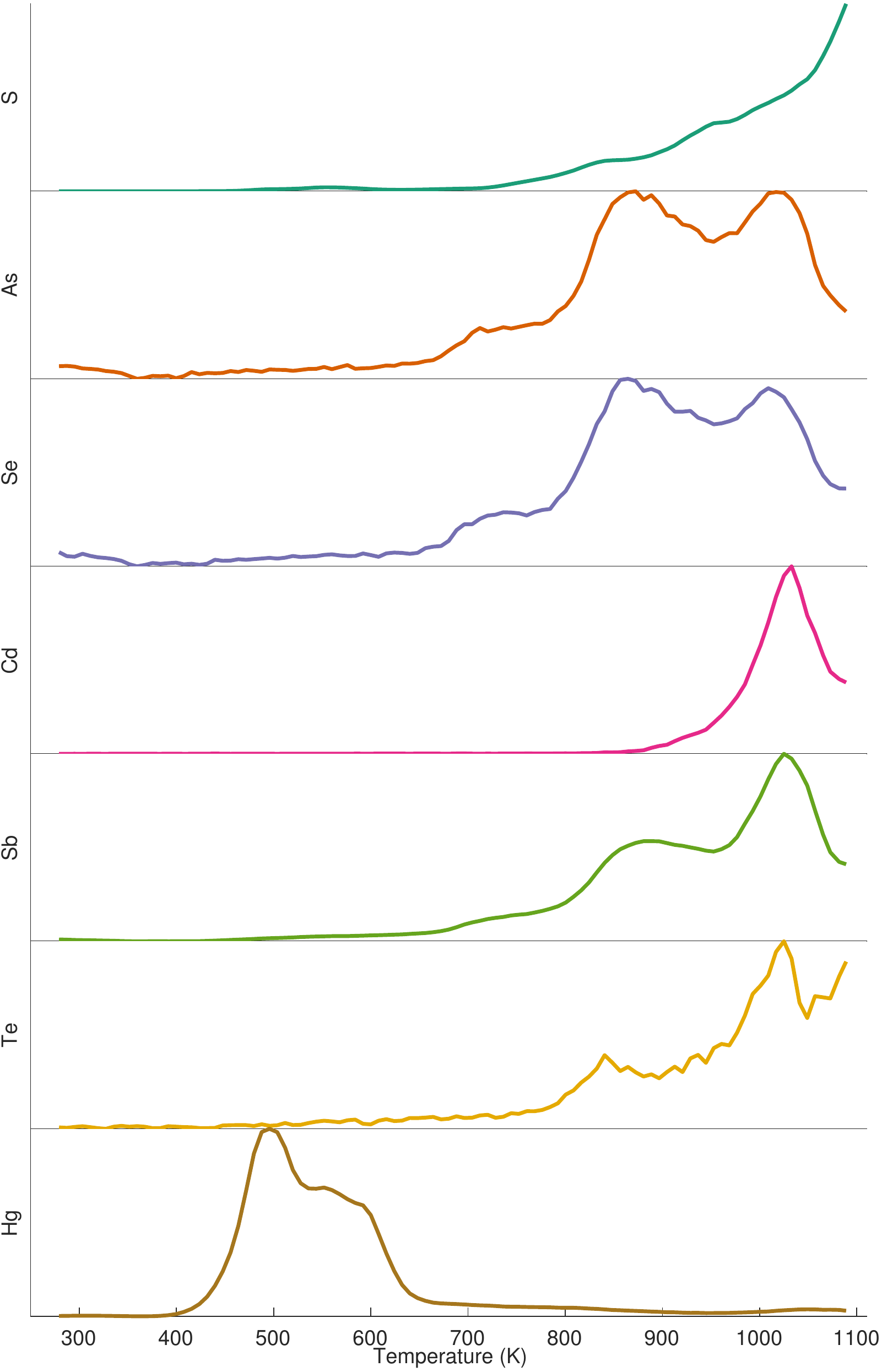}
    \caption{Allende: S shows a small peak at 560 K, then begins a linear increase in release above 650 K. As and Se show similar release profiles with two peaks at 870 and 1010 K.  Cd, Sb, and Te also have a peak at 1025 K, and Sb an additional peak at 880 K.  Hg peaks at 495 K.}
    \label{fig:ab-allende}
\end{subfigure}
    \caption{\AbunCaption}
    \end{figure}
\end{landscape}    
    
\clearpage   
 
\begin{landscape} 
\begin{figure}[tb]\ContinuedFloat
\begin{subfigure}[t]{\SubAbunWidth\hsize}
\centering    
\includegraphics[width=\AbunWidth\linewidth]{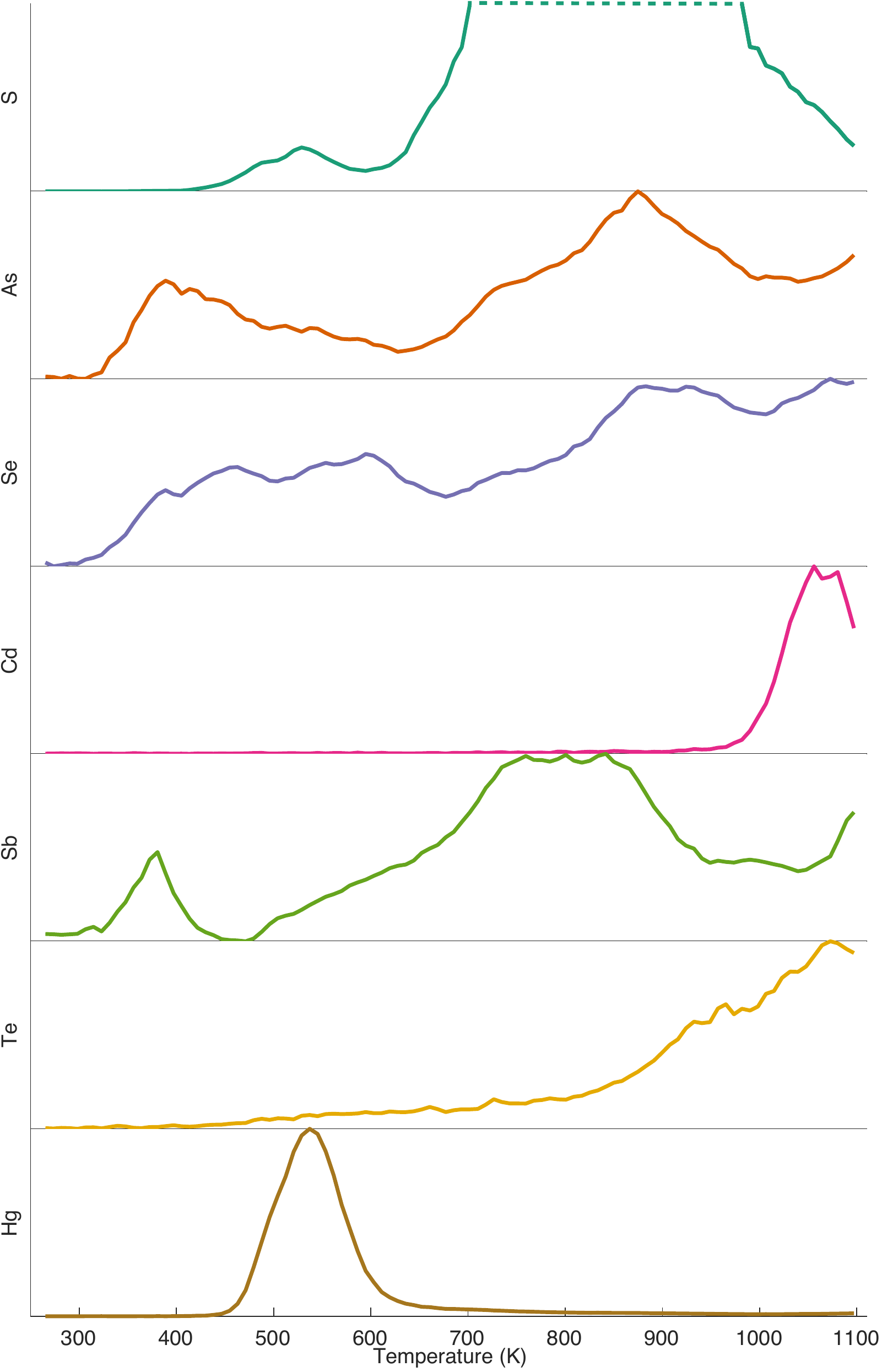}
    \caption{Grosnaja: S has a small peak at 530 K; the dashed line represents detector saturation between 700 and 980 K.  As and Se have  peaks at 380 and 885 K. Cd has one large peak at 1055 K. Sb has a small peak at 380 K and a peak between 760 and 840 K. Te continuously releases, peaking at 1075 K. Hg has a peak at 535 K.}
    \label{fig:ab-grosnaja}
\end{subfigure}   
\hfill
\begin{subfigure}[t]{\SubAbunWidth\hsize}
\centering    
\includegraphics[width=\AbunWidth\linewidth]{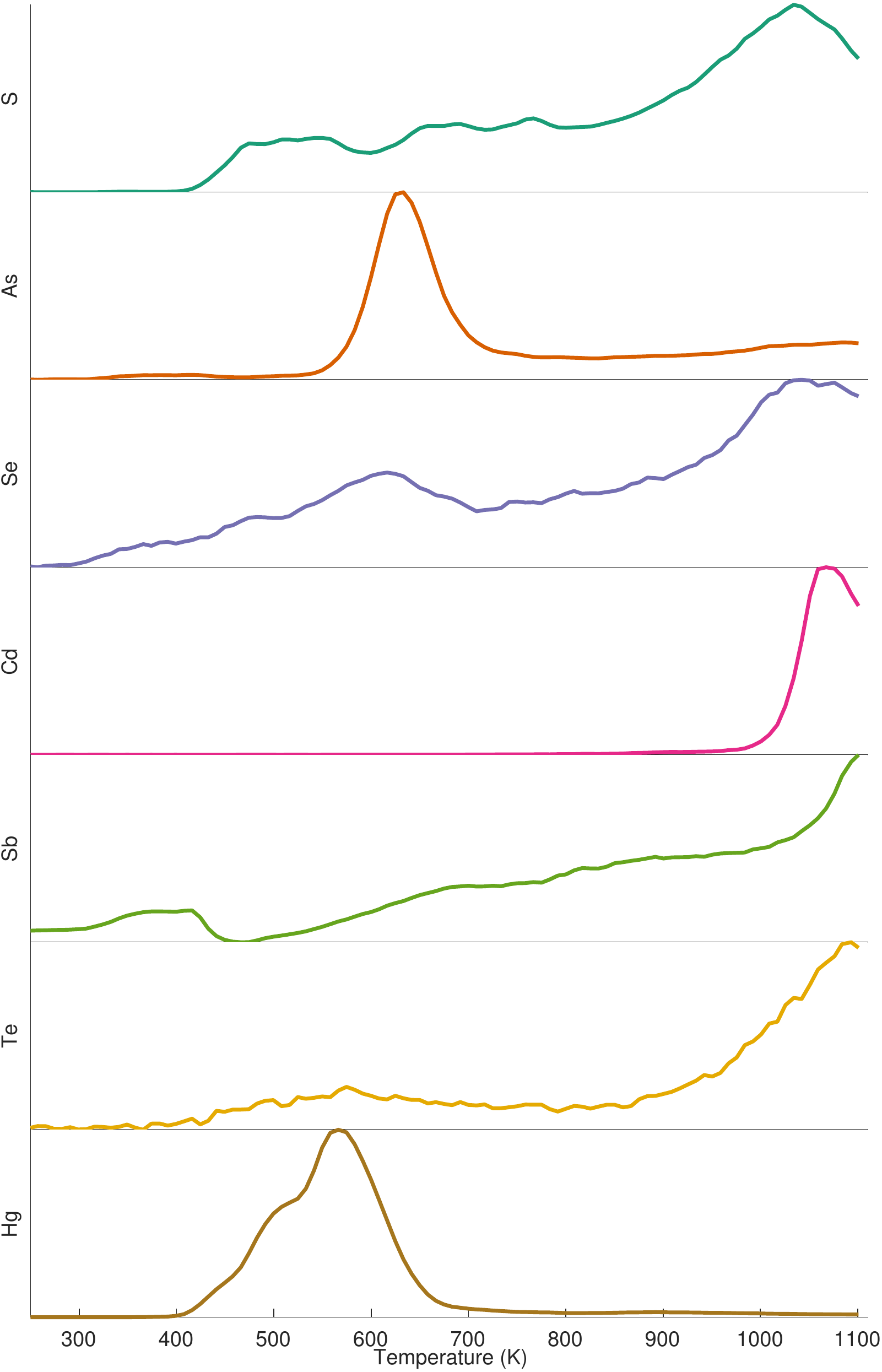}
    \caption{Mokoia: S begins releasing at 415 K with a large peak at 1035 K. As has a large peak in release at 635 K. Se peaks at 615 and 1042 K. Cd has one peak at 1065; Sb follows a similar profile with a minor peak at 415 K, then increasing to a maximum at 1100 K. Te has a small peak at 570 K, then a large one at 1095 K. Hg peaks at 565 K, close to the first Te peak.}
    \label{fig:ab-mokoia}
\end{subfigure}
    \caption{\AbunCaption}
    \end{figure}
\end{landscape}    

\clearpage

\begin{landscape} 
\begin{figure}[tb]\ContinuedFloat
\begin{subfigure}[t]{\SubAbunWidth\hsize}
\centering    
\includegraphics[width=\AbunWidth\linewidth]{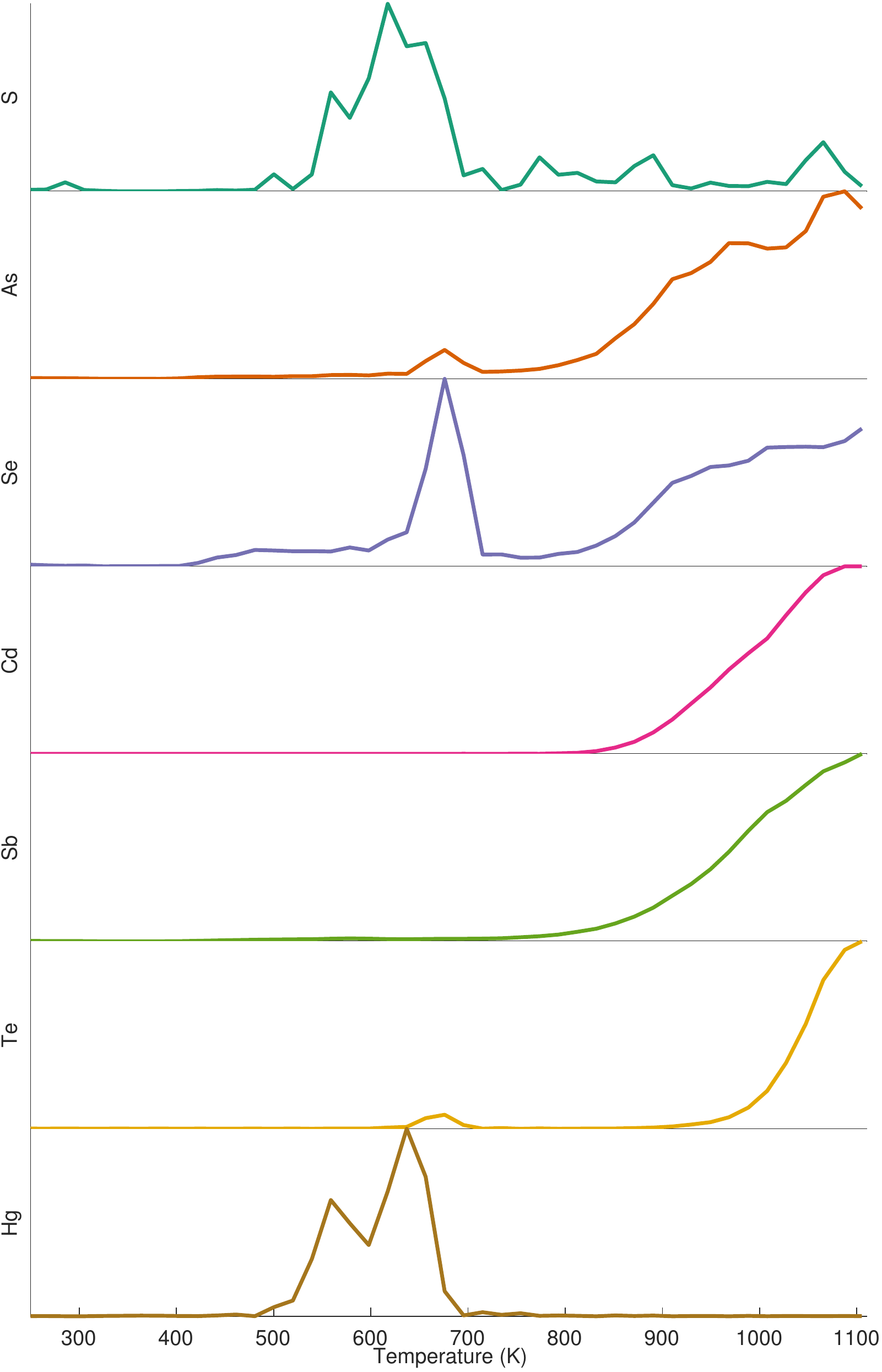}
    \caption{Vigarano: this experiment was run at a lower resolution than for other samples. S shows maximum release at 615 K. As and Se peak at 675 K and continue releasing from the sample; As has a large peak at 1085 K. Cd and Sb show flat release profiles until 815 and 715 K, respectively. Te has a small peak at 675 K, then increases after 900 K. Hg has two peaks at 560 and 640 K.}
    \label{fig:ab-vigarano}
\end{subfigure}   
\hfill
\begin{subfigure}[t]{\SubAbunWidth\hsize}
\centering    
\includegraphics[width=\AbunWidth\linewidth]{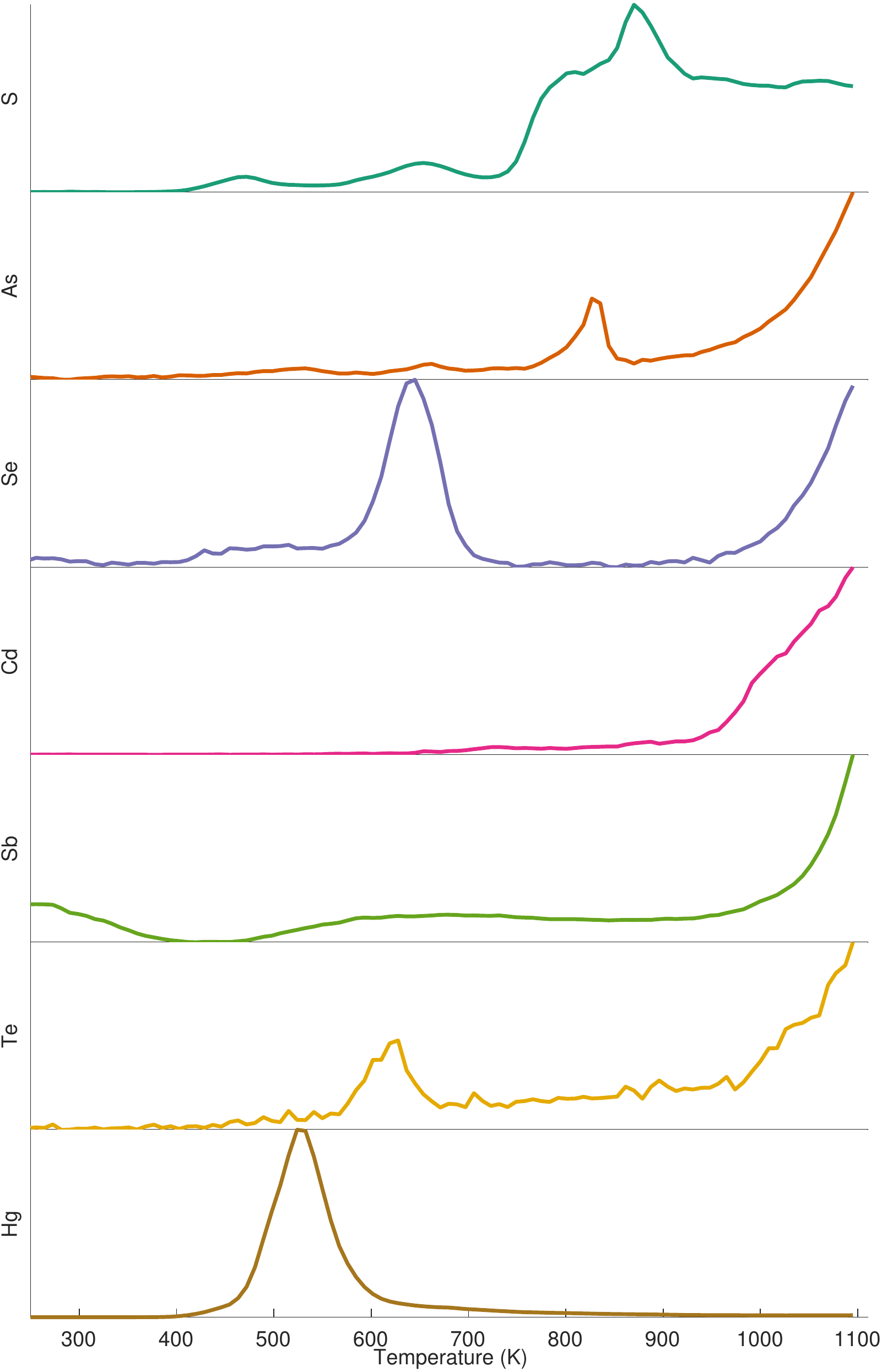}
    \caption{Kainsaz: S release has peaks at 465, 655, and 870 K. 
    As shows small peaks at 530, 660, and 825, with counts increasing after 875 K. 
    Se has a large peak at 645 K, then increases release above 900 K. 
    Cd shows a flat release profile until 650 K, then has a sharp increase in counts released at 930 K. 
    Sb counts initially decrease until 420 K, then increase after 930 K. 
    Te has a small peak at 625 K, then increases after 740 K. 
    Hg peaks at 535 K. 
    }
    \label{fig:ab-kainsaz}
\end{subfigure}
    \caption{\AbunCaption}
 \end{figure}
\end{landscape}

\clearpage

\begin{landscape} 
\begin{figure}[tb]\ContinuedFloat
\begin{subfigure}[t]{\SubAbunWidth\hsize}
\centering    
\includegraphics[width=\AbunWidth\linewidth]{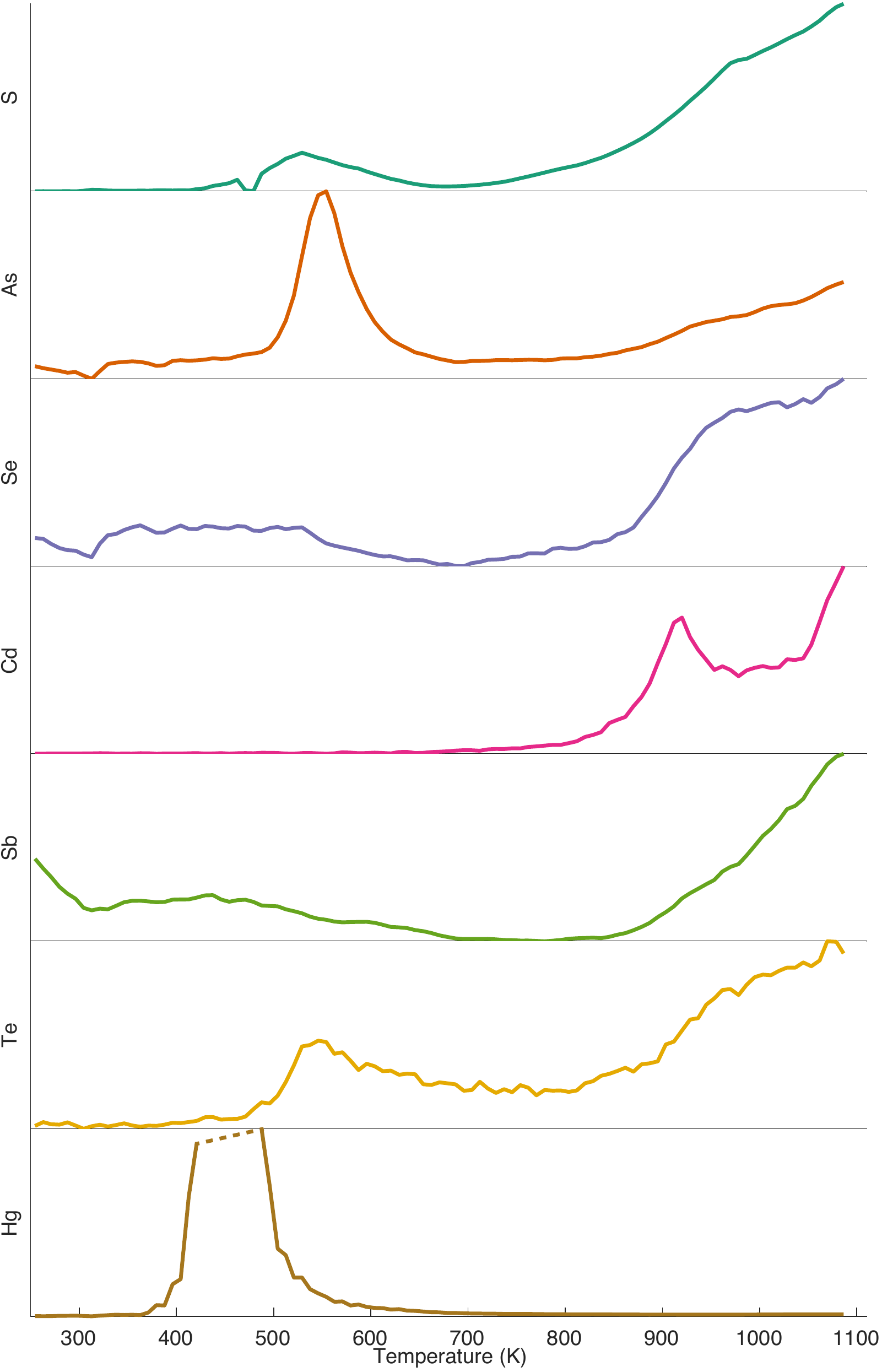}
    \caption{Ornans: S release peaks at 520 K then increases after 675 K. As peaks at 550 K, then increases after 775. Se shows increased release between 310 and 530 K, then increases above 695 K. Cd release peaks at 920 K and increases above 975. Sb release decreases until 780 K, then increases up to 1085 K. Te peaks at 545 and 1070 K. The dashed line for Hg represents detector saturation between 420 and 485 K. 
    }
     \label{fig:abunornans}
\end{subfigure}   
\hfill
\begin{subfigure}[t]{\SubAbunWidth\hsize}
\centering    
\includegraphics[width=\AbunWidth\linewidth]{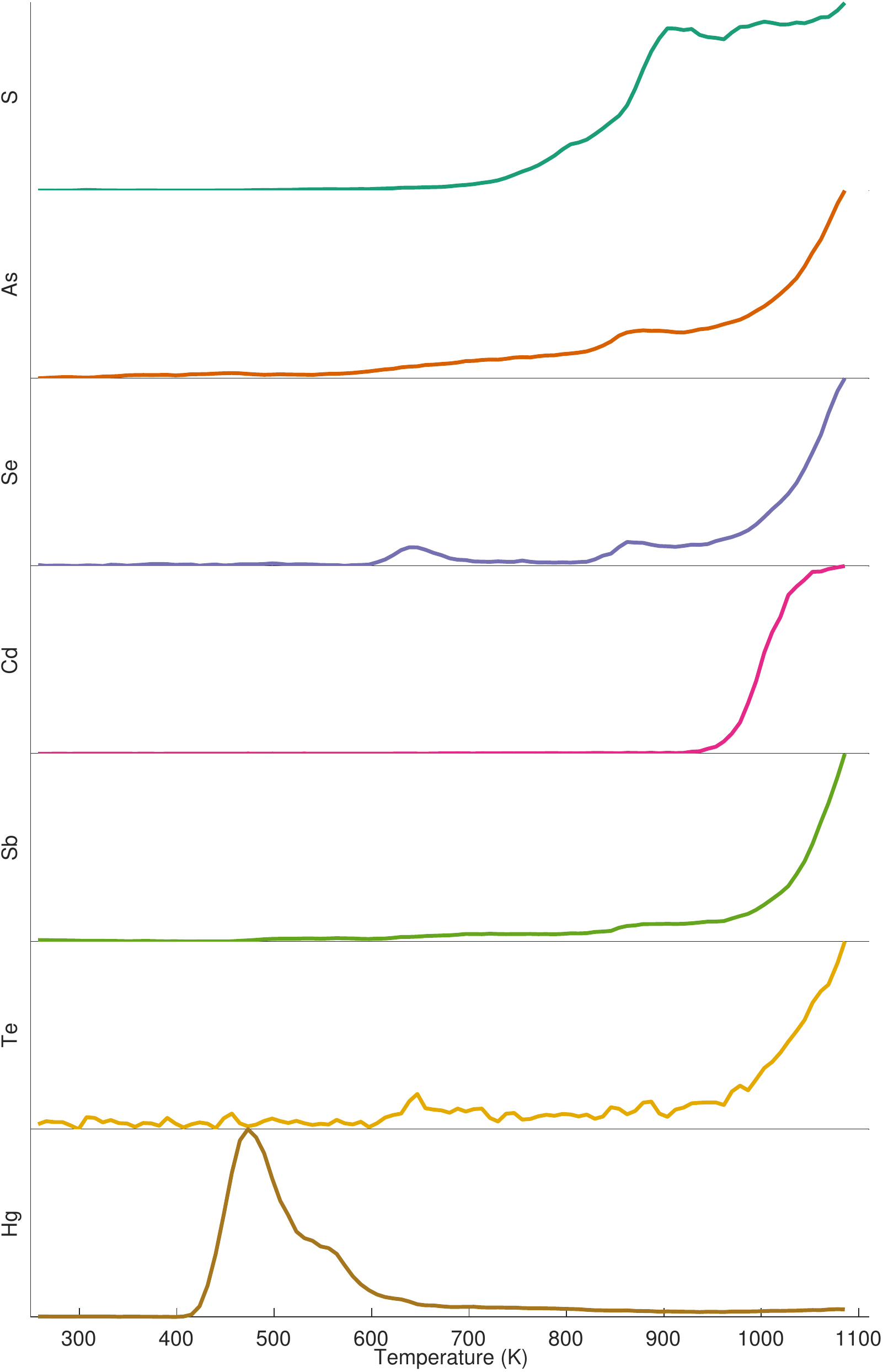}
    \caption{Isna: S, As, Se, Cd, Sb, and Te all have relatively flat release profiles, eventually increasing. S counts increase above 600 K with a peak at 910 K. As counts increase above 540 K with a small peak at 875. Se has small peaks at 640 and 870 K, then increases above 910 K. Cd counts increase above 925 K; Sb above 845 K; Te above 945 K. Hg has a peak at 470 K. 
    }
    \label{fig:abunisna}
\end{subfigure}
    \caption{\AbunCaption}
 \end{figure}
\end{landscape}

\newcommand{\PerCaption}{Percentage lost of each element versus temperature, for each meteorite sample. Some elements were below detection level, or not appreciably lost.}
\newcommand{\PerWidth}{0.9}

\begin{figure}[tb]
\centering 
\begin{subfigure}[t]{1.0\hsize}
\centering    
\includegraphics[width=\PerWidth\linewidth]{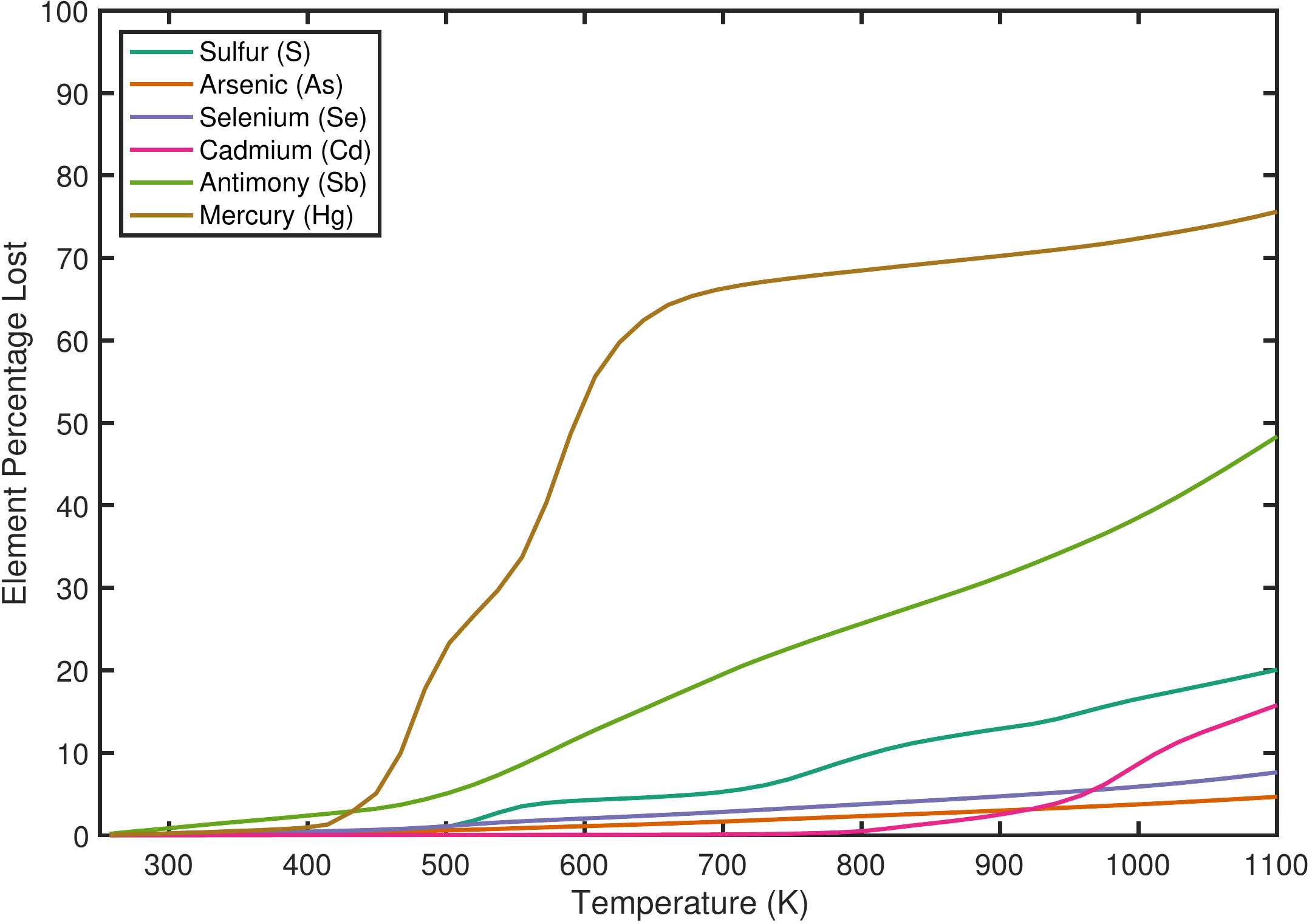}
    \caption{Orgueil. Orgueil shows over 60\% loss of Hg by 650 K, with Sb depleted by $> 16$\% at the same temperature.  The maximum Hg lost is 75\%; Sb, 48\%; S, 20\%; Cd, 16\%; and the remaining two elements lose less than 4\%.
    }
\end{subfigure}   

\begin{subfigure}[t]{1.0\hsize}
\centering    
\includegraphics[width=\PerWidth\linewidth]{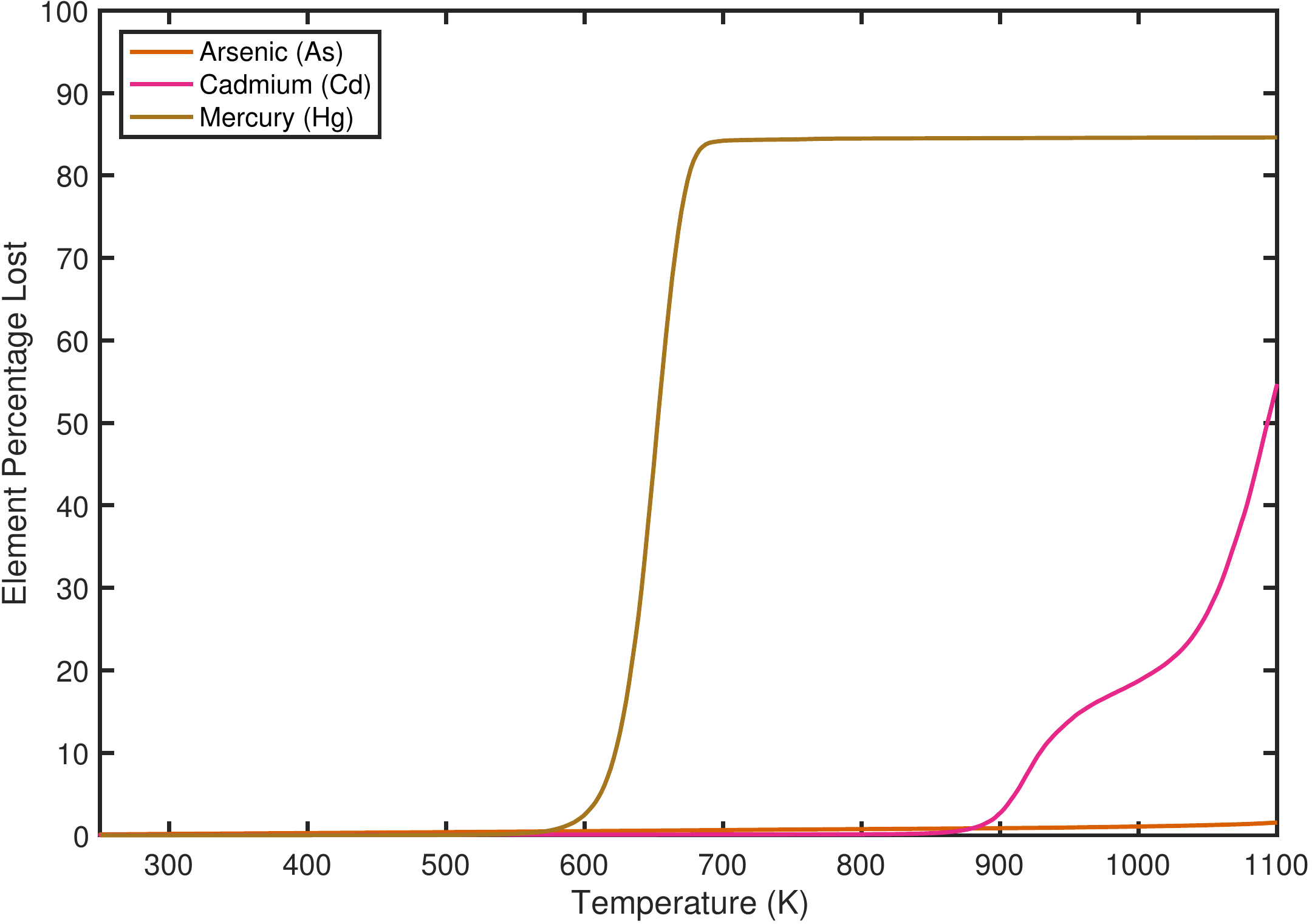}
    \caption{Murchison. Murchison loses 80\% of its Hg by 675 K, and 85\% by 1100. The host phase of Cd begins release Cd at 900 K, evidenced by 54\% loss of Cd at 1100 K.  A scant 1.5\% of As is lost by 1100 K.}
\end{subfigure}
\caption{\PerCaption}
 \label{fig:percentlost}
\end{figure}

\clearpage

\begin{figure}[tb]\ContinuedFloat
\centering 
\begin{subfigure}[t]{1.0\hsize}
\centering    
\includegraphics[width=\PerWidth\linewidth]{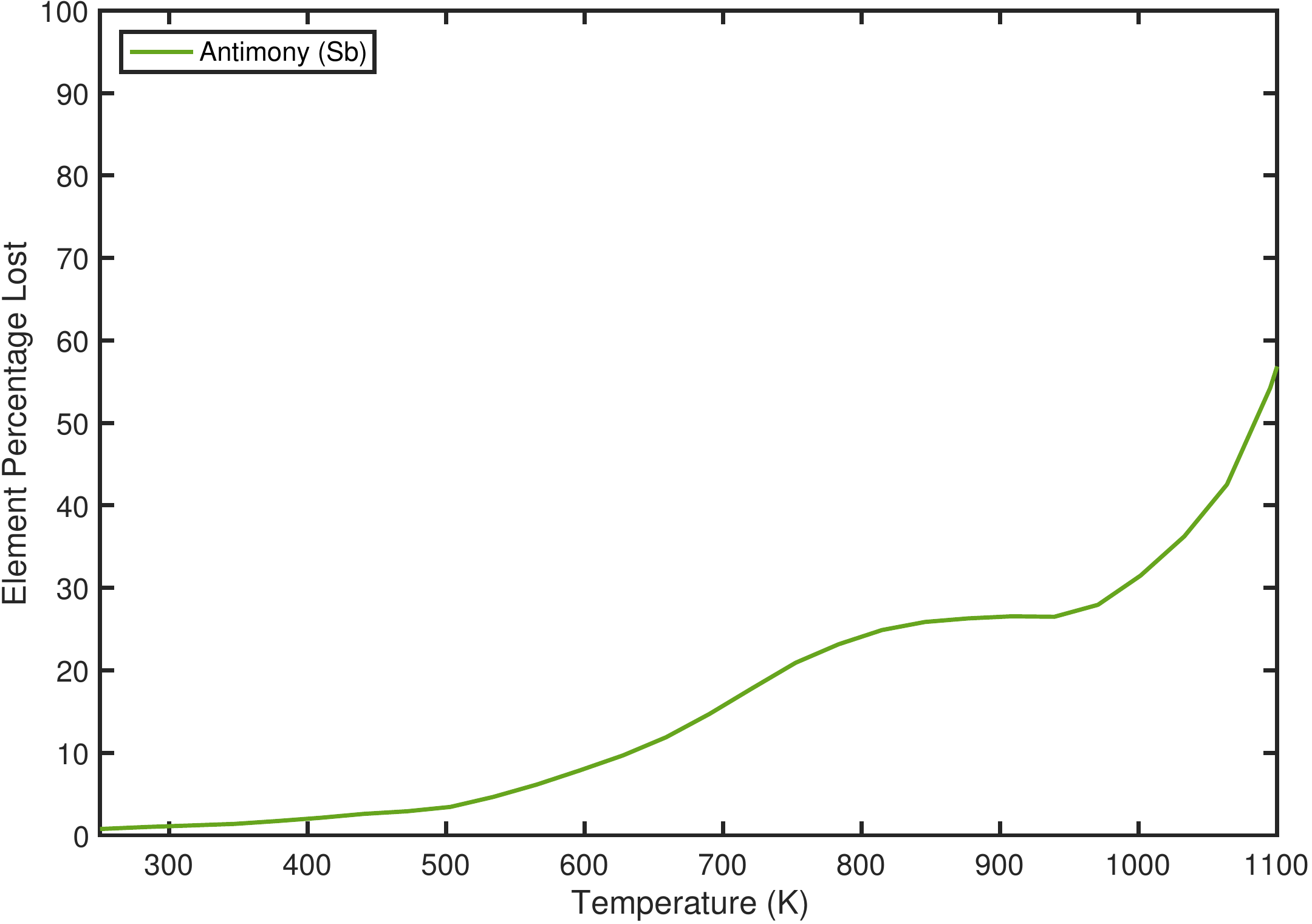}
    \caption{Murray. Only Sb was detected as lost from Murray, with release above 5\% at 650 K up to 54\% at 1100 K.}
\end{subfigure}   

\begin{subfigure}[t]{1.0\hsize}
\centering    
\includegraphics[width=\PerWidth\linewidth]{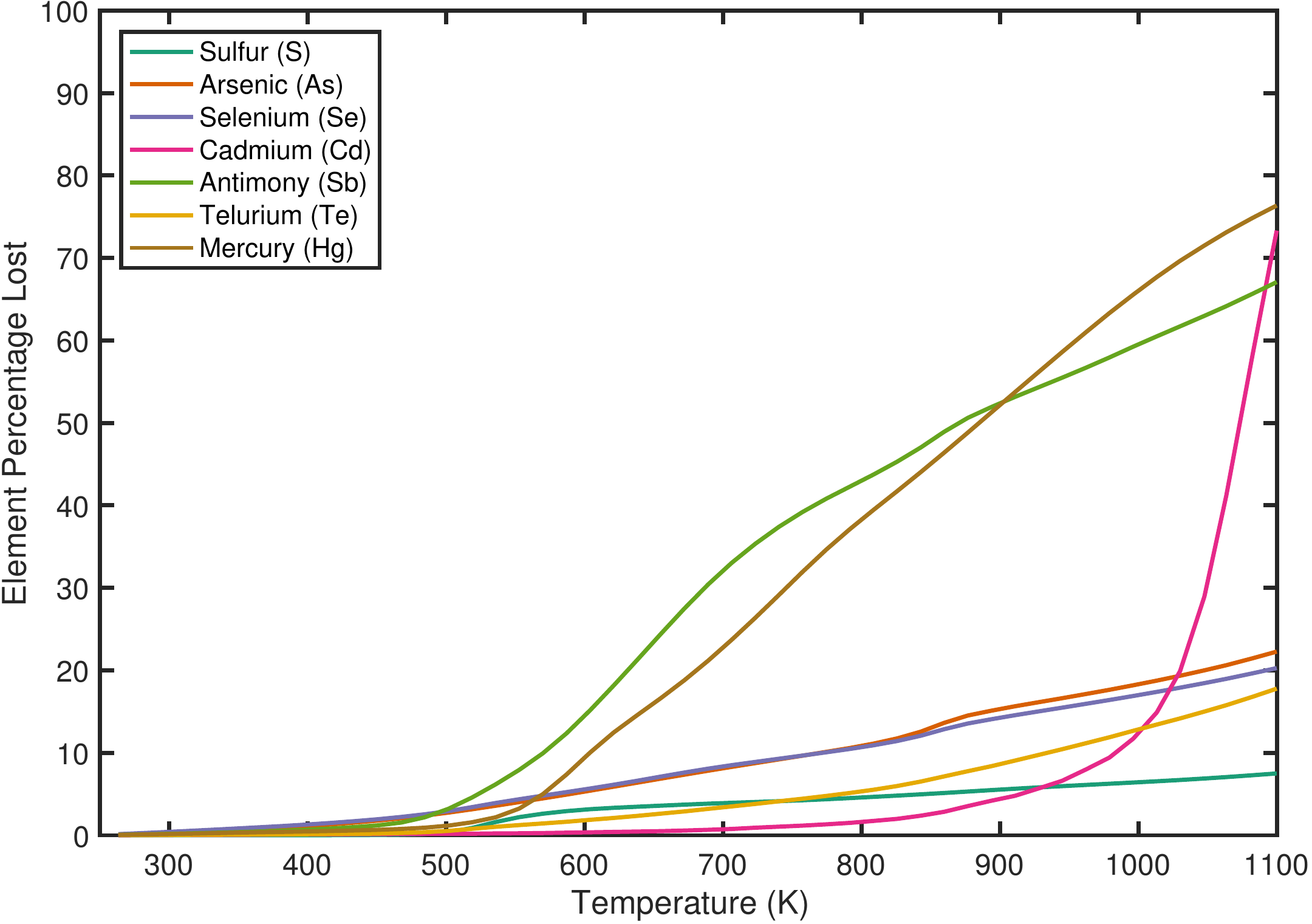}
    \caption{Nogoya. All seven measured elements were detected as released from Nogoya. Substantive loss of both Sb and Hg begins at 500 K, and continues until 67\% and 76\% of each element, respectively, are lost by 1100 K. Cd begins a rapid rate of loss above 1000 K, losing 73\%.  Loss rates for As, Se, Te, and S remain flat and less than 25\%.}
\end{subfigure}
\caption{\PerCaption}
    \end{figure}

\clearpage

\begin{figure}[tb]\ContinuedFloat
\centering 
\begin{subfigure}[t]{1.0\hsize}
\centering    
\includegraphics[width=\PerWidth\linewidth]{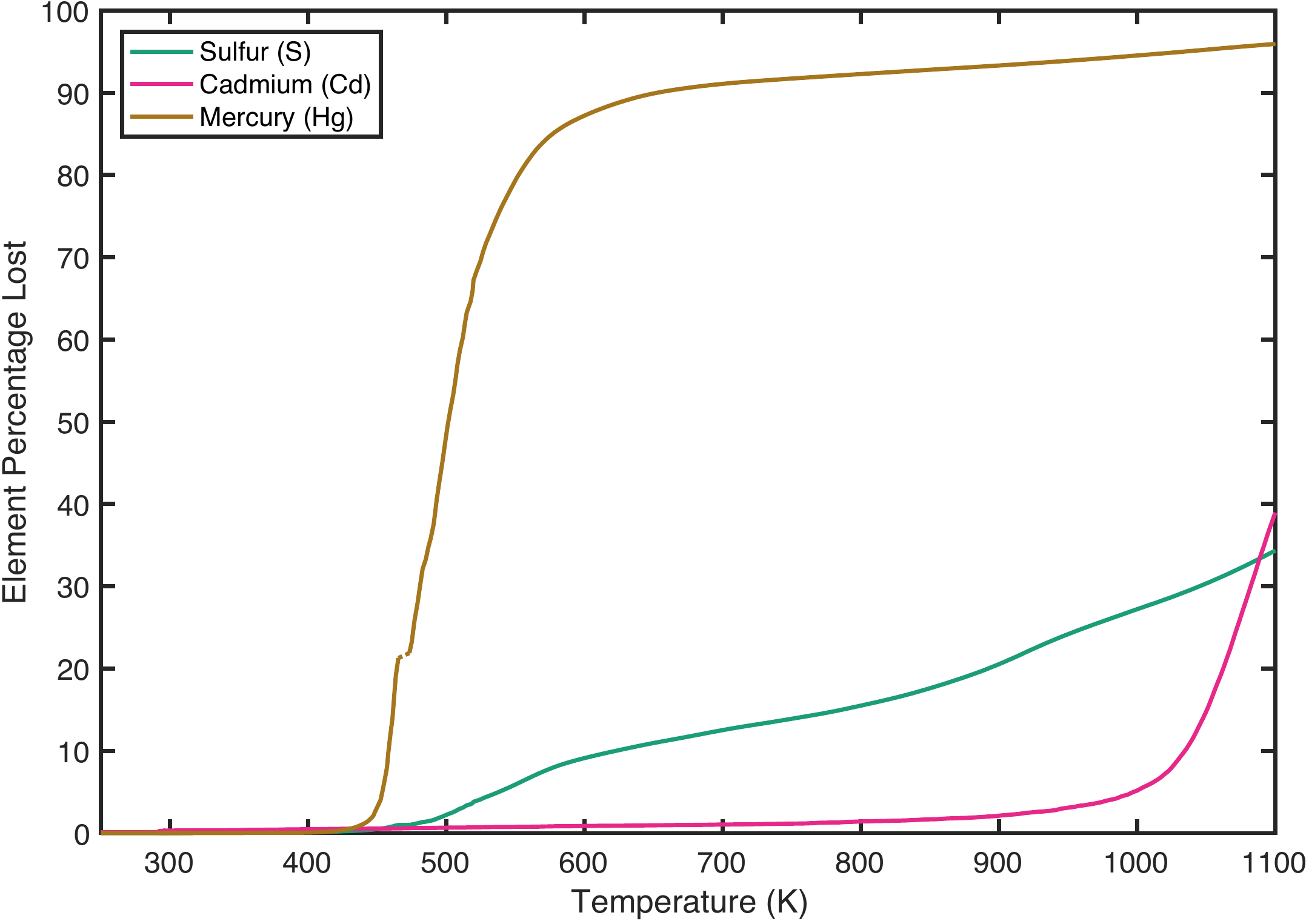}
    \caption{Cold Bokkeveld. Rapid loss of Hg begins at 430 K; the dashed line between 465 and 475 K shows where the detector saturated owing to high count rates for Hg.  Ultimately, 96\% of Hg is lost.  The loss rate of Cd is flat until $\sim1000$ K, and 38\% of Cd is lost by 1100.  S is lost gradually between 450 and 1100 K, with 34\% lost overall.}
\end{subfigure}   

\begin{subfigure}[t]{1.0\hsize}
\centering    
\includegraphics[width=\PerWidth\linewidth]{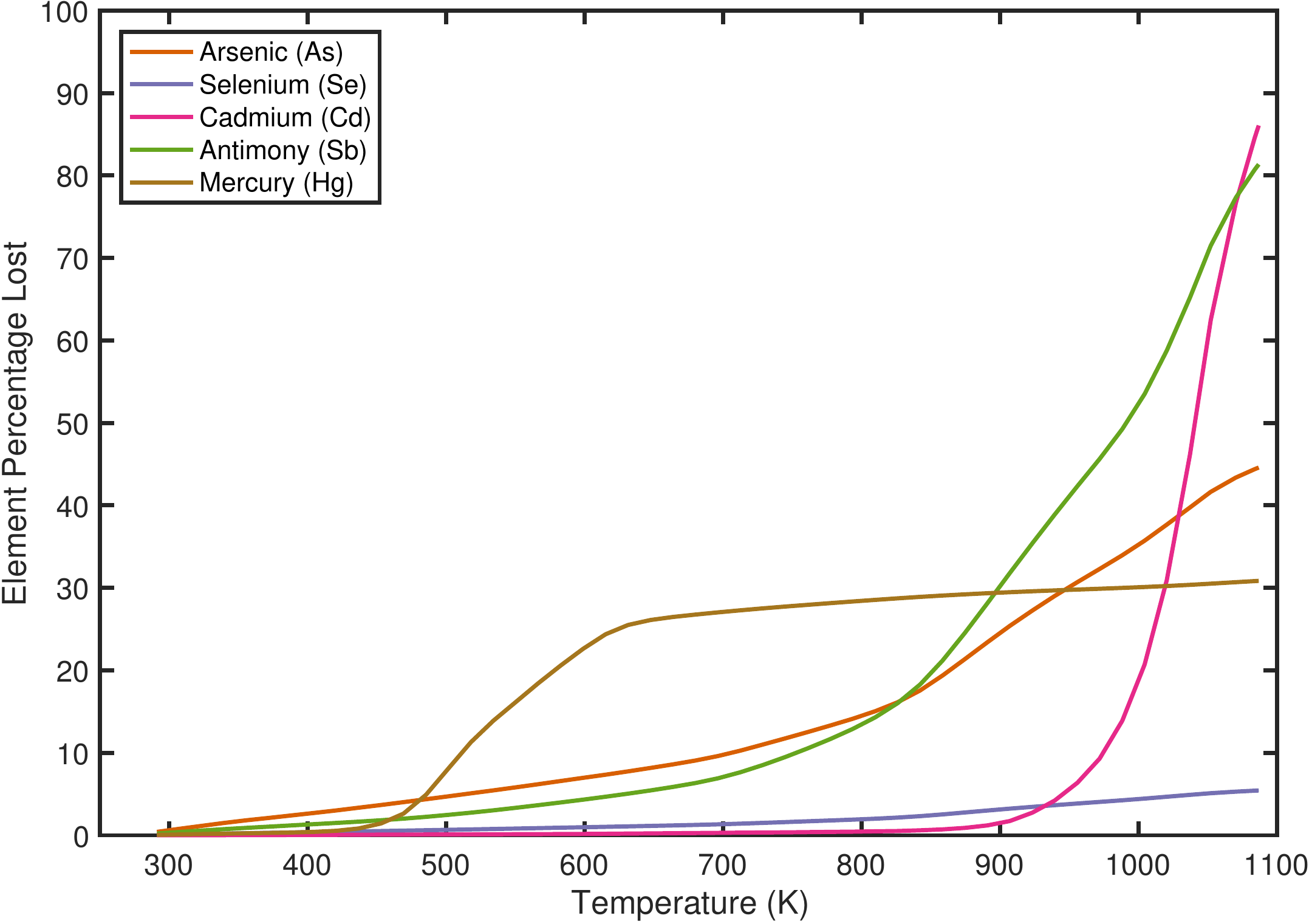}
    \caption{Allende. Hg reaches its maximum loss at $\sim625$ K, with 30\% lost overall.  As and Sb increase their loss rates above 750 K, losing 45 and 80\%, respectively.  Cd begins rapid release at 900 K and loses 86\% of the element.  About 5\% of Se is lost.}
\end{subfigure}
\caption{\PerCaption}
    \end{figure}

\clearpage

\begin{figure}[tb]\ContinuedFloat
\centering 
\begin{subfigure}[t]{1.0\hsize}
\centering    
\includegraphics[width=\PerWidth\linewidth]{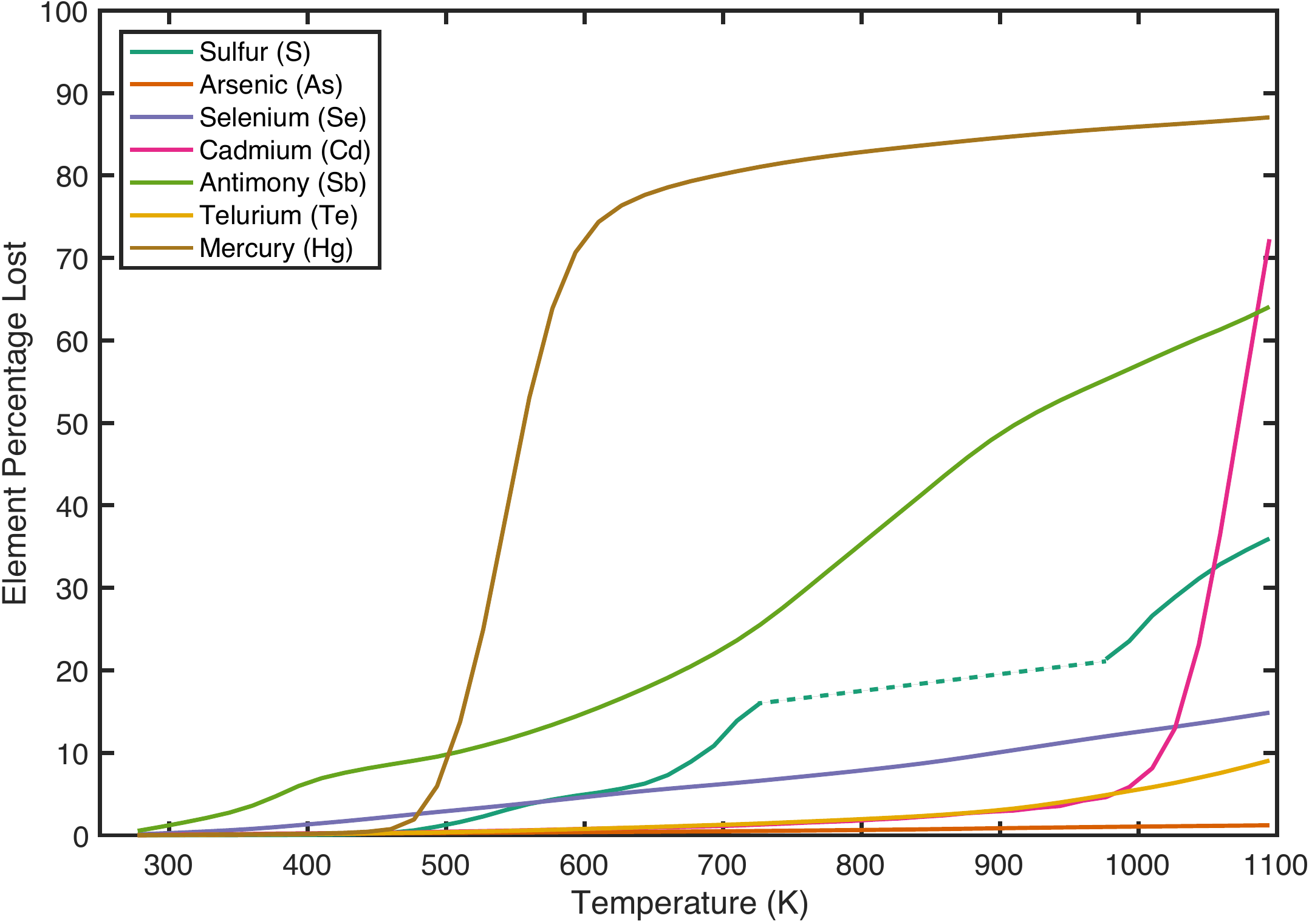}
    \caption{Grosnaja. Rapid loss of Hg begins at 450 K, with $>$75\% of Hg lost at 650 K, and a maximum of 87\% list.  Sb is lost gradually throughout the entire temperature range, up to 65\%.  
    The jagged shape of S is due to saturation in the detector when measuring this element, with 64\% lost by 1090 K.  
        A rapid increase in Cd loss begins at 1025 K, up to 72\%.  
        Less than 15\% of Se, Te, and As are lost by 1090 K.
        }
\end{subfigure}   

\begin{subfigure}[t]{1.0\hsize}
\centering    
\includegraphics[width=\PerWidth\linewidth]{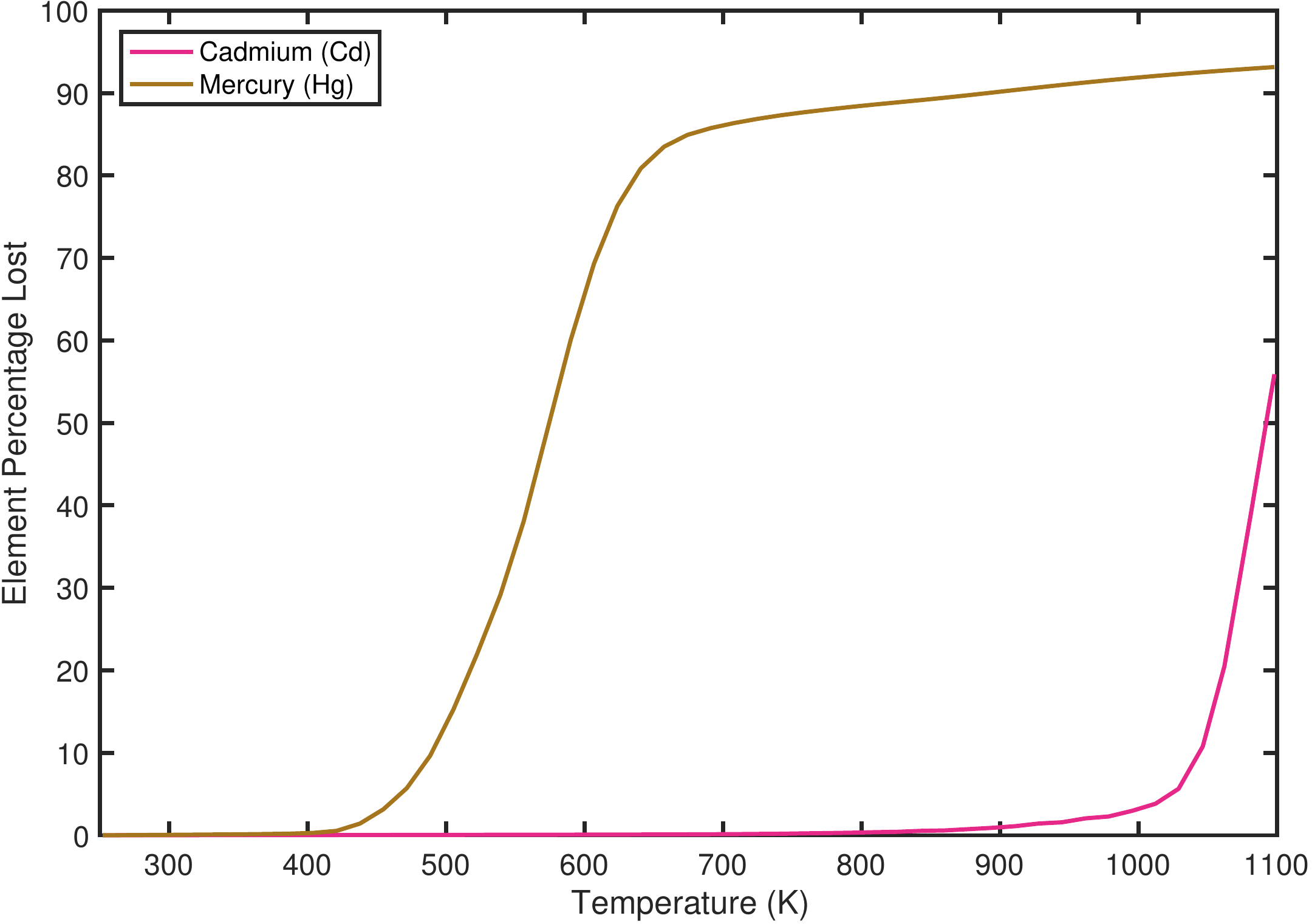}
    \caption{Mokoia. Rapid loss of Hg begins at 475 K, with 80\% lost by 650 K, and 93\% lost overall. Cd loss is low until 1025 K, increasing to 56\% by 1100 K.}
\end{subfigure}
\caption{\PerCaption}
    \end{figure}

\clearpage

\begin{figure}[tb]\ContinuedFloat
\centering 
\begin{subfigure}[t]{1.0\hsize}
\centering    
\includegraphics[width=\PerWidth\linewidth]{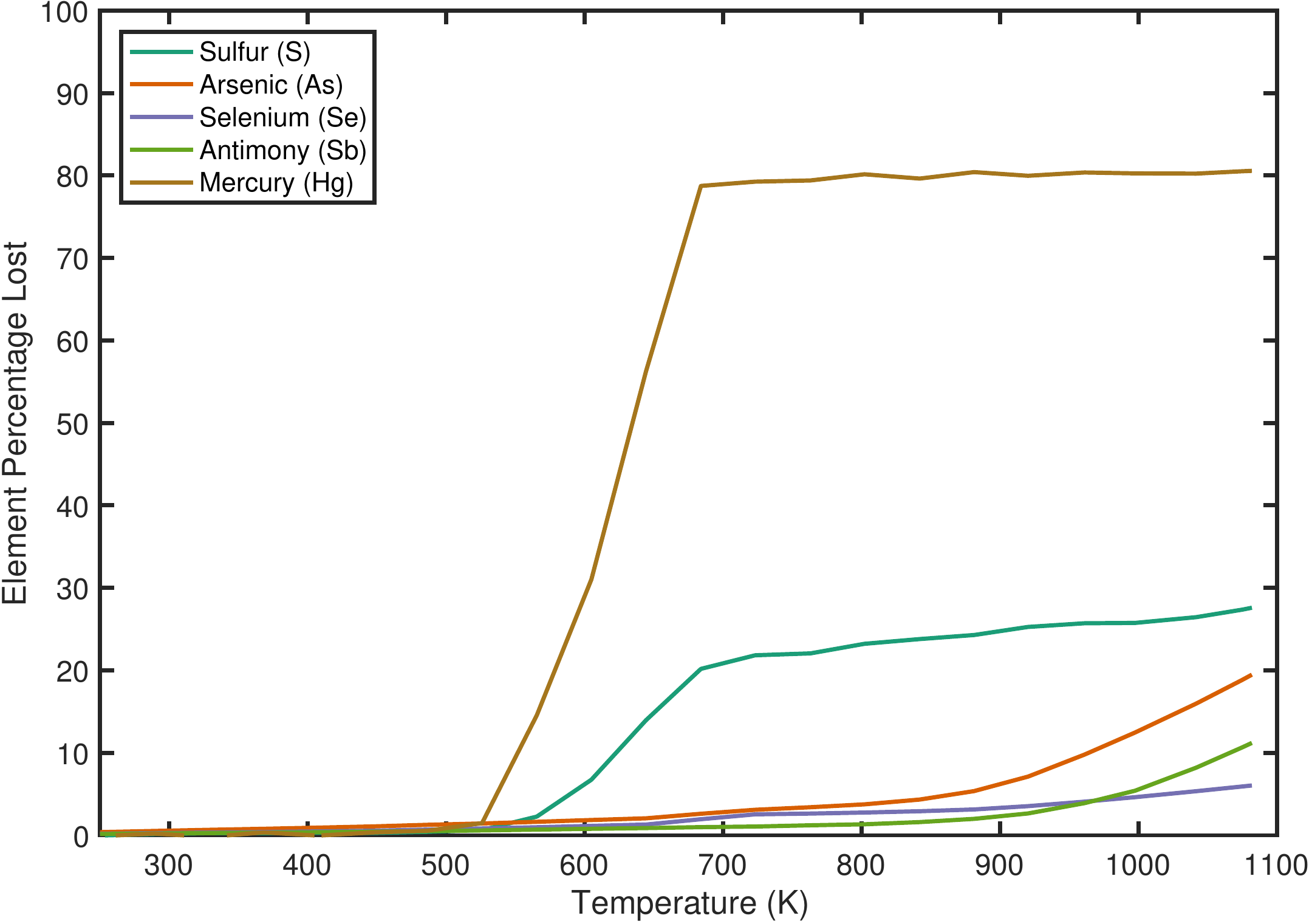}
    \caption{Vigarano. By 675 K, Vigarano has lost the majority of its Hg (80\%), with the greatest rate of S loss occurring between 600-675 K, then increasing more linearly up to overall loss of 28\%.  As, Se, and Sb are lost more gradually, up to 19, 6, and 11\%, respectively.}
\end{subfigure}   

\begin{subfigure}[t]{1.0\hsize}
\centering    
\includegraphics[width=\PerWidth\linewidth]{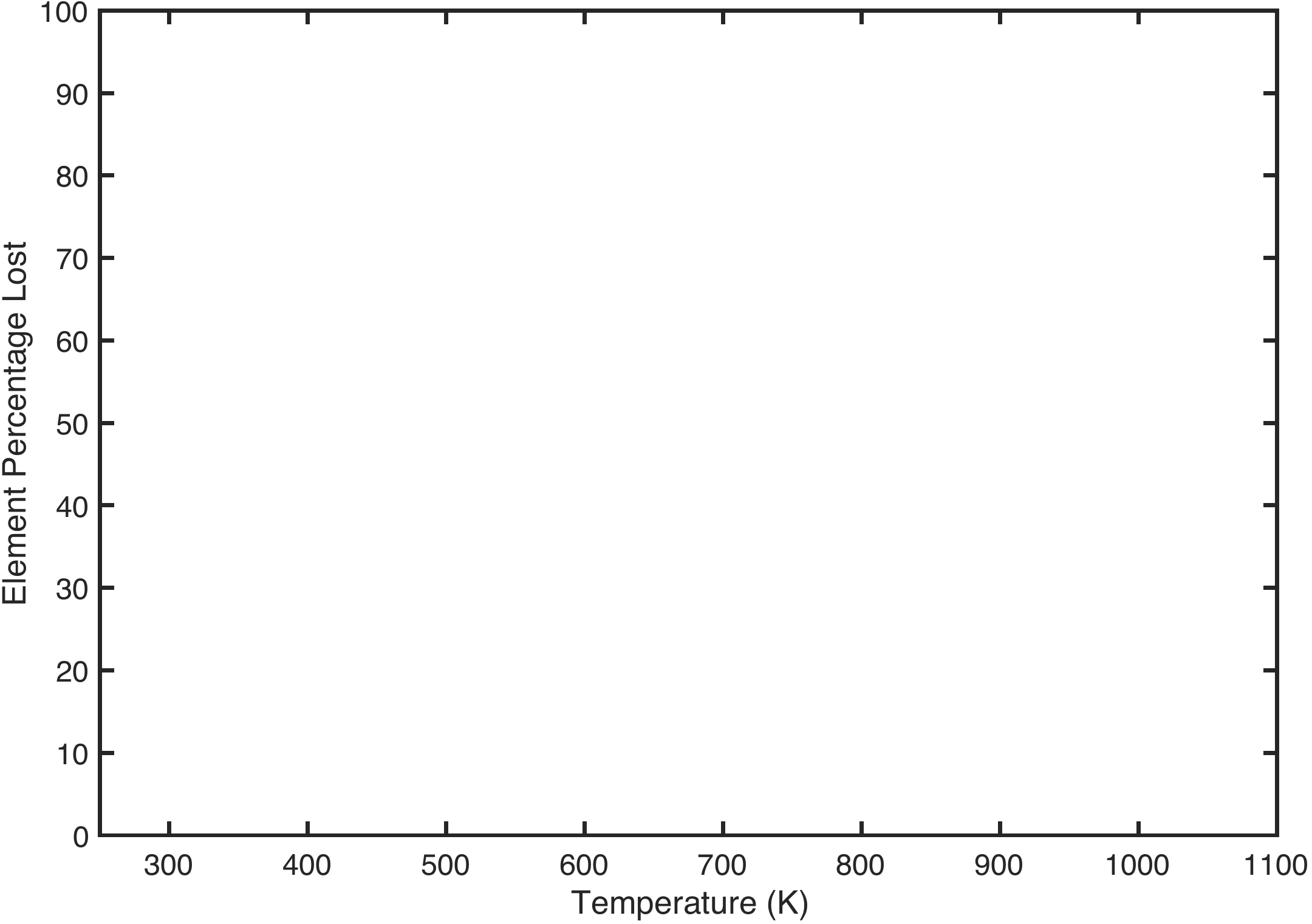}
    \caption{Kainsaz. Because the heating experiments consumed the entire available sample mass, the total elemental loss as a result of heating was not measured for Kainsaz; thus we could not calculate the percentages of elements lost.}
\end{subfigure}
\caption{\PerCaption}
    \end{figure}
    
\clearpage   
 
\begin{figure}[tb]\ContinuedFloat
\centering 
\begin{subfigure}[t]{1.0\hsize}
\centering    
\includegraphics[width=\PerWidth\linewidth]{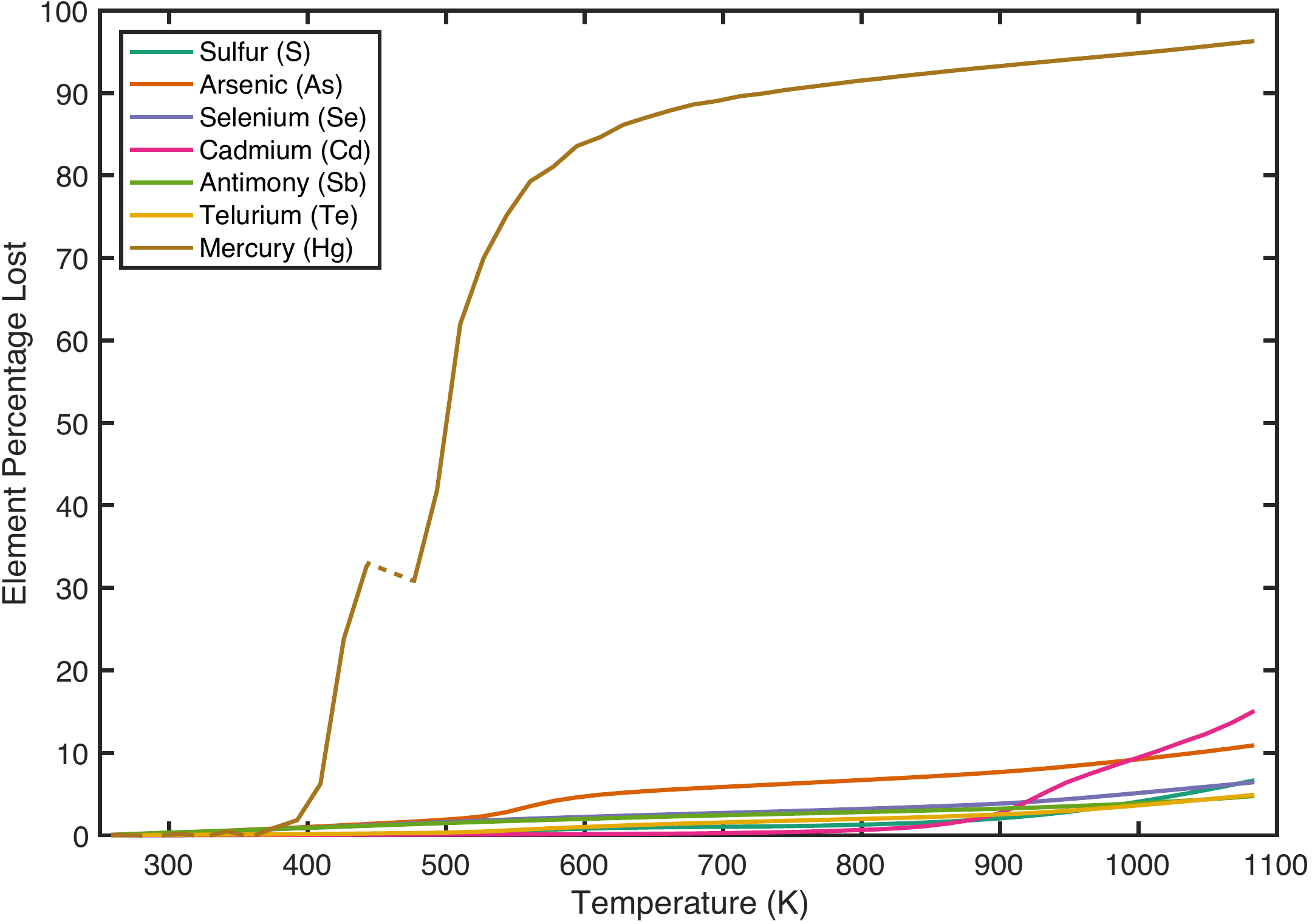}
    \caption{Ornans. The discontinuities in Hg loss between 440--475 K, represented by a dashed line, are due to saturation in the detector during element measurement.  Eighty-eight percent of Hg is lost by 650 K, and 96\% by 1080 K.  Other elements show gradual losses up to 5-15\% by 1100 K.}
\end{subfigure}   

\begin{subfigure}[t]{1.0\hsize}
\centering    
\includegraphics[width=\PerWidth\linewidth]{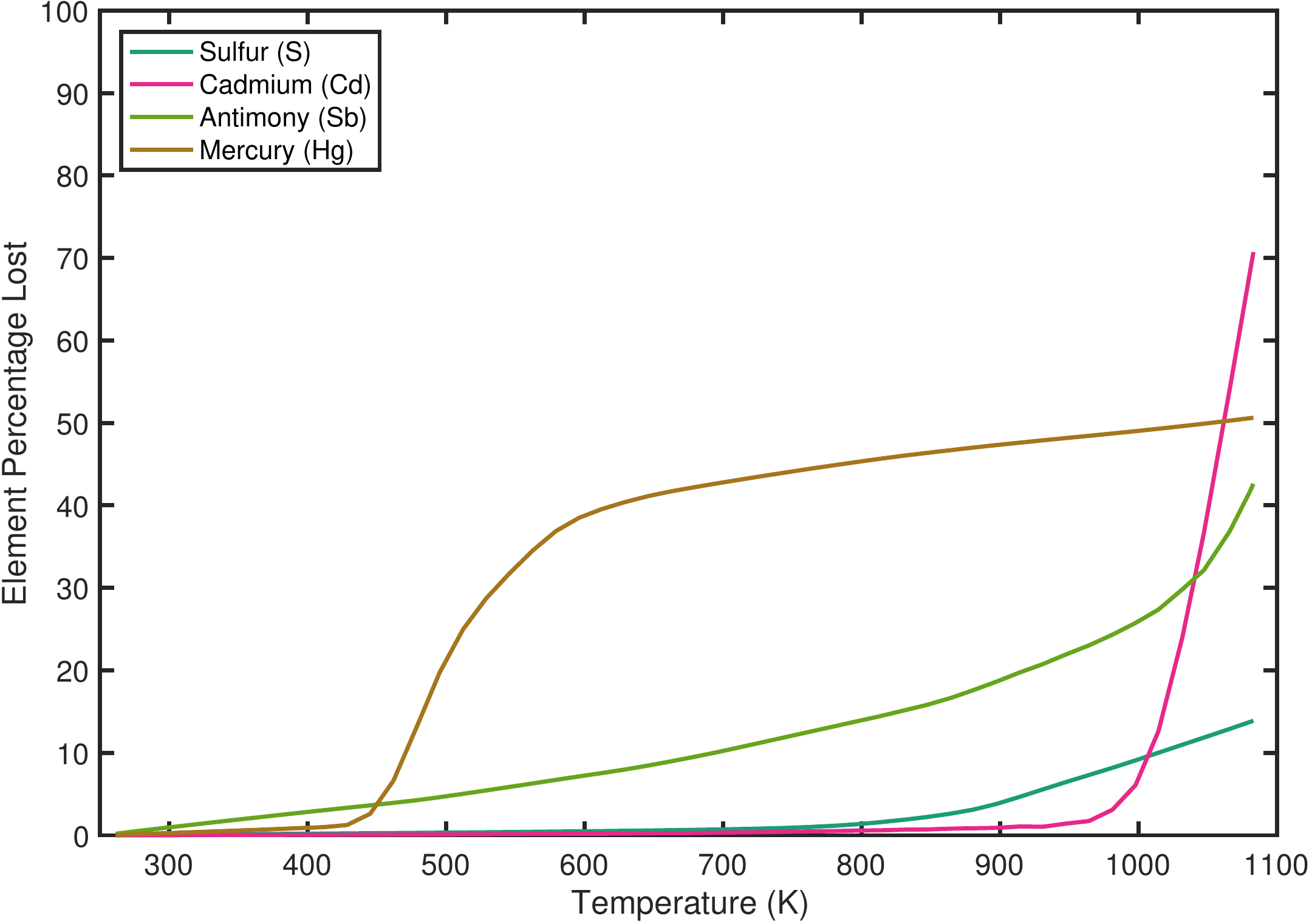}
    \caption{Isna. Rapid loss of Hg begins at 450 K; by 650 K, 41\% is lost, and 50\% by 1080 K.  Cd loss increases at 950 K, up to 71\%.  Sb and S are lost gradually, up to 43 and 14\%, respectively.}
\end{subfigure}
\caption{\PerCaption}
    \end{figure}

\begin{landscape}    
\begin{figure}[tb]
\centering    
\begin{subfigure}[t]{0.8\hsize}    
\includegraphics[width=\columnwidth]{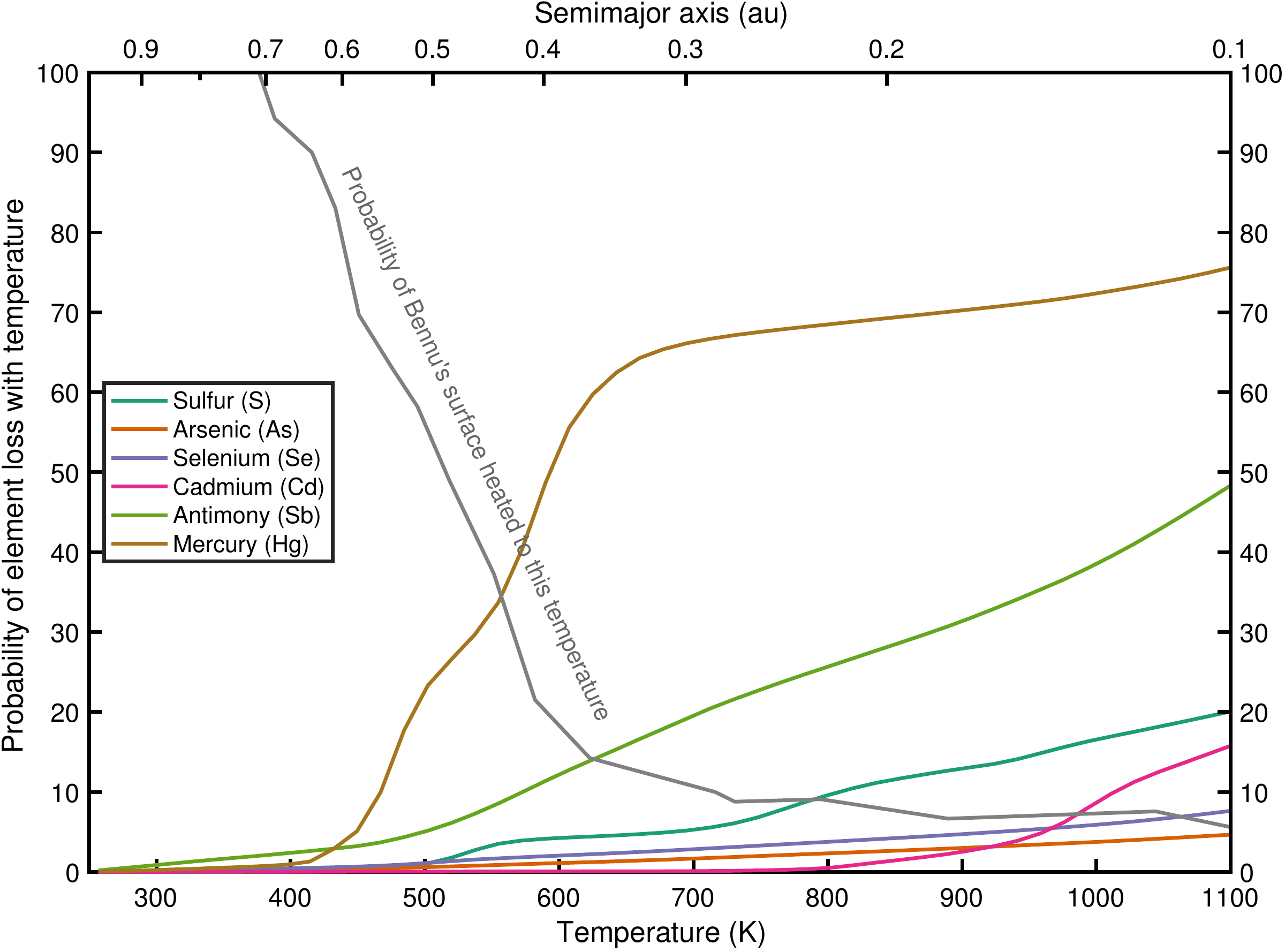}
\caption{Orgueil}
\end{subfigure}
\end{figure}
\end{landscape}

\clearpage

\begin{landscape}    
\begin{figure}[tb]\ContinuedFloat
\centering    
\begin{subfigure}[t]{0.8\hsize}    
\includegraphics[width=\columnwidth]{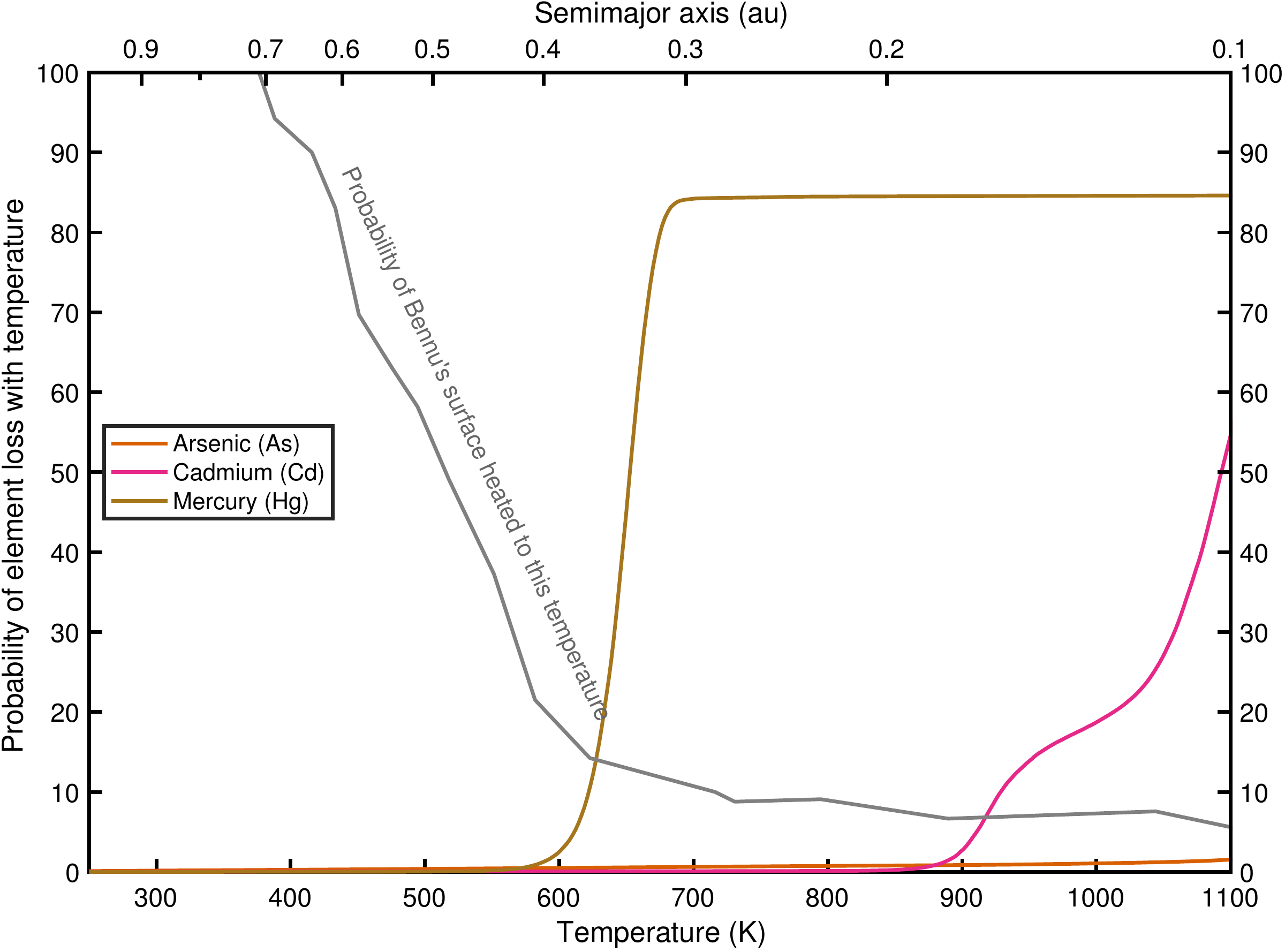}
\caption{Murchison}
\end{subfigure}
\end{figure}
\end{landscape}

\clearpage

\begin{landscape}
\begin{figure}[tb]\ContinuedFloat
\centering    
\begin{subfigure}[t]{0.8\hsize}    
\includegraphics[width=\columnwidth]{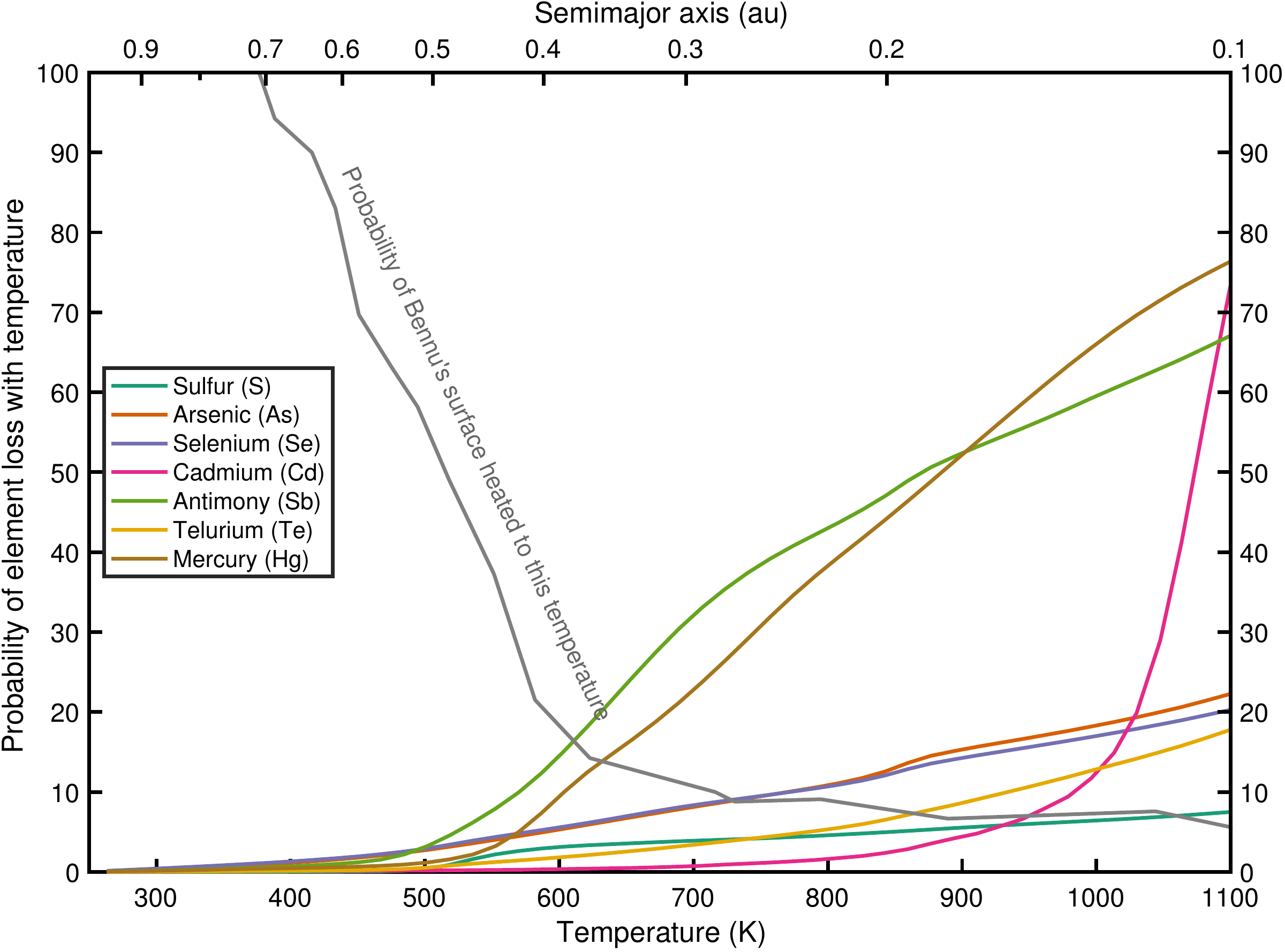}
\caption{Nogoya}
\end{subfigure}
\caption{Element percentage loss versus temperature for Orgueil, Murchison, and Nogoya plotted with the probability of Bennu's subsolar surface having experienced past heating versus temperature \citep{Delbo2011}.  If Bennu has experienced heating up to 625 K we would expect loss of $\sim$15\% of Hg in returned Bennu regolith similar to a Murchison composition.  The top x-axis perihelion distances were calculated assuming a blackbody at the equilibrium temperatures along the bottom x-axs.  Adapted from \citet{Delbo2011}, with permission.}
\label{fig:delbo_murchison}
\end{figure}
\end{landscape}

\end{document}